\documentclass[twocolumn]{aastex62}
\usepackage{array} 
\usepackage{amsmath}
\usepackage{wasysym}

\newenvironment{tightcenter}{%
	\setlength\topsep{0pt}
	\setlength\parskip{0pt}
	\begin{center}
}{%
 	\end{center}
}

\newcommand\chandra{{\it Chandra}}

\newcommand\ciao{CIAO}
\newcommand\xspec{XSPEC}
\newcommand\caldb{CALDB}

\shorttitle{Variability and Proper Motion in Centaurus\,A}
\shortauthors{Snios et al.}

\begin{document}

\title{Variability and Proper Motion of X-ray Knots in the Jet of Centaurus\,A}

\author{Bradford Snios} 
\affil{Harvard-Smithsonian Center for Astrophysics, 60 Garden St, Cambridge, MA 02138, USA}
\author{Sarka Wykes}
\affil{Independent Researcher}
\affil{Harvard-Smithsonian Center for Astrophysics, 60 Garden Street, Cambridge, MA 02138, USA}
\author{Paul E. J. Nulsen} 
\affil{Harvard-Smithsonian Center for Astrophysics, 60 Garden Street, Cambridge, MA 02138, USA}
\affil{ICRAR, University of Western Australia, 35 Stirling Hwy, Crawley, WA 6009, Australia}
\author{Ralph P. Kraft}
\affil{Harvard-Smithsonian Center for Astrophysics, 60 Garden Street, Cambridge, MA 02138, USA}
\author{Eileen T. Meyer}
\affil{Department of Physics, University of Maryland Baltimore County, Baltimore, MD 21250, USA}
\author{Mark Birkinshaw}
\affil{H. H. Wills Physics Laboratory, University of Bristol, Tyndall Ave, Bristol BS8 1TL, UK}
\author{Diana M. Worrall}
\affil{H. H. Wills Physics Laboratory, University of Bristol, Tyndall Ave, Bristol BS8 1TL, UK}
\author{Martin J. Hardcastle}
\affil{School of Physics, Astronomy and Mathematics, University of Hertfordshire, College Lane, Hatfield, Hertfordshire AL10 9AB, UK}
\author{Elke Roediger}
\affil{E.A. Milne Centre for Astrophysics, School of Mathematics and Physical Sciences, University of Hull, Hull, HU6 7RX, UK}
\author{William R. Forman}
\affil{Harvard-Smithsonian Center for Astrophysics, 60 Garden Street, Cambridge, MA 02138, USA}
\author{Christine Jones}
\affil{Harvard-Smithsonian Center for Astrophysics, 60 Garden Street, Cambridge, MA 02138, USA}

\begin{abstract}
\noindent We report results from \chandra{} observations analyzed for evidence of variability and proper motion in the X-ray jet of Centaurus\,A. Using data spanning 15 yr, collective proper motion of \mbox{$11.3\pm3.3 \rm\ mas\ yr^{-1}$}, or \mbox{$0.68\pm0.20c$}, is detected for the fainter X-ray knots and other substructure present within the jet. The three brightest knots (AX1A, AX1C, and BX2) are found to be stationary to an upper limit of  $0.10c$. Brightness variations up to 27\% are detected for several X-ray knots in the jet. For the fading knots, BX2 and AX1C, the changes in spectral slope expected to accompany synchrotron cooling are not found, ruling it out and placing upper limits of $\simeq 80\rm\ \mu G$ for each of their magnetic field strengths. Adiabatic expansion can account for the observed decreases in brightness. Constraints on models for the origin of the knots are established. Jet plasma overrunning an obstacle is favored  as the generator of stationary knots, while moving knots are likely produced either by internal differences in jet speed or the late stages of jet interaction with nebular or cloud material.
\end{abstract}

\keywords{galaxies: active -- galaxies: individual (Centaurus A) -- galaxies: jets -- X-rays: galaxies}

%%%%%%%%%%%%%%%%%%%%%%%%%%%
\section{Introduction}
\label{sect:intro}

Kiloparsec-scale X-ray jets have been recognized as a hallmark of extragalactic radio sources for several decades \citep[e.g.,][]{Feigelson1981, Harris1997, Turner1997, Hardcastle2001, Worrall2001}. Radiation from lower-power, i.e.,~Fanaroff-Riley class~I (FR~I; \citealp{Fanaroff1974}), jets is argued to be X-ray synchrotron in nature based on the monotonic decrease in spectral intensity with increasing frequency, high-levels of linear polarization, and the lack of the gamma-ray emission expected if inverse Compton is the dominant emission process \citep[e.g.,][]{Hardcastle2001, Hardcastle2006, Kraft2002, Marshall2002, Worrall2010, Meyer2014, Gentry2015, Meyer2015, Breiding2017}. The radiative lifetime of the X-ray-emitting electrons is only tens of years, assuming there are equipartition magnetic fields, which implies that particle acceleration must be local. This is useful when considering the likely scenario that the properties of extragalactic jet flows are dominated by their interactions with components of small volume-filling fractions \citep[e.g.,][]{Aluzas2012, Wykes2015}.

Proper motion studies provide us with a means of directly observing projected jet velocities or, at a minimum, the pattern speed of the jet. However, direct observations of jet motion on kiloparsec scales are scarce, with the optical study by \cite{Meyer2017} providing the largest sample to date. Their work has shown that FR~I jets have a slowly increasing jet speed up to a distance of 100~pc from the nucleus, and decelerate on larger scales. There are a few reports of apparent jet acceleration on parsec scales, measured using very long baseline interferometry (VLBI), in other FR~I sources \citep[e.g.,][]{Cotton1999, Lister2013, Nagai2014, Hada2016, Boccardi2017} and also in Centaurus\,A \citep[e.g.,][]{Muller2014} -- the target of our work. It remains to be determined whether in some sources this genuinely is an increase in speed on the small scales, as one could be assessing speed in different jet layers if the jet has a spine-sheath structure. Gradual jet deceleration progressively over 0.1--15~kpc scales has been predicted and is now found in several FR~I radio sources \citep[e.g.,][]{Bicknell1994, Laing1999, Laing2002, Canvin2004, Canvin2005, Meyer2013, Perucho2014b}.

\begin{figure}
	\begin{tightcenter}
	\includegraphics[width=0.47\textwidth]{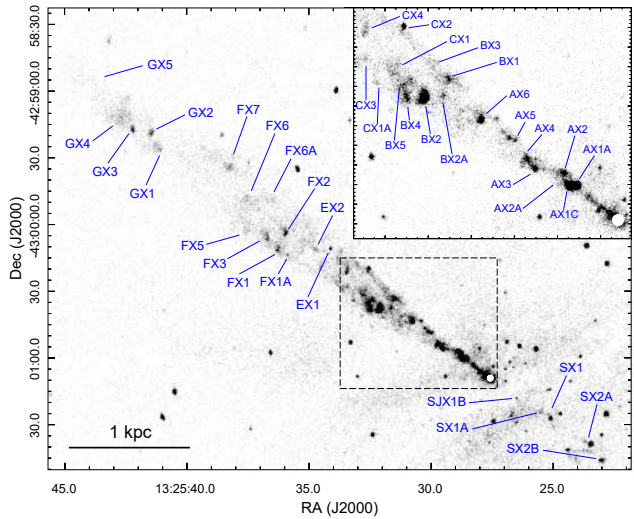}
	\end{tightcenter}
    \caption{0.9--2.0 keV, exposure-corrected 
    		\chandra{} image of the Centaurus A jet and counterjet. ACIS-S 
		observations listed in Table~\ref{table:obs} were co-added for the image.
		Pixel size is 0.492\arcsec, and the image has been smoothed with 
		an 8 pixel RMS Gaussian filter (inset smoothed with 5 pixel rms 
		Gaussian). Shown are X-ray knot identifications, and the inset 
		(top right) displays the knots of the ``inner jet." 
		The active galactic nuclei (AGN) has been masked in the image.}
	\label{fig:cena}
\end{figure}

Due to its proximity, at $3.8 \pm 0.1\rm\ Mpc$ \citep{Harris2010}, Centaurus\,A (Cen\,A) is the only synchrotron jet where the \chandra{} imaging instrument can probe the distance traveled by electrons before synchrotron losses remove them from the X-ray band. Indeed, observations of Cen\,A have revealed a wealth of detail within the jet, reinforcing its X-ray synchrotron nature and placing constraints on the particle acceleration and the origin of the patchy surface brightness enhancements (knots) in the jet (Figure~\ref{fig:cena}; \citealp{Kraft2002, Hardcastle2003, Kataoka2006, Goodger2010}). In studying how the X-ray spectra of the knots vary with distance from the jet axis, \cite{Worrall2008} demonstrated that the knots reside at a range of off-axis angles rather than being confined to a shear layer between faster and slower flows. \cite{Hardcastle2001} and \cite{Goodger2010} detected radio proper motion for 3 of the 40 knots identified in the jet. Those moving knots showed comparatively little X-ray emission, indicating that high-energy electron acceleration is less efficient in these structures than at those with zero apparent motion.

While the origin of the radio and X-ray bright knots in Cen\,A has been investigated in depth by \cite{Goodger2010}, no single model for knot formation -- including adiabatic compression, impulsive particle acceleration, collisions with stationary objects, and recollimation shocks -- adequately explains all the observed features. Recollimation shocks provide a poor description of the knots in Cen\,A, as they do not extend to the full width of the jet \citep{Tingay2009, Goodger2010}. Pure compressions in the synchrotron-emitting plasma have been argued against elsewhere \citep{Hardcastle2003}. Cen\,A shows a widening region in the jet at $\sim$\,260~pc (projected) near the most upstream and brightest X-ray knots AX1A and AX1C, suggesting that the abrupt increase in jet diameter associated with those knots are shocks at relatively large obstructions in the jet. \cite{Goodger2010} put forward collisions with stationary objects as the most likely scenario for the origin of the majority of the knots in Cen\,A, and they proposed compressions in the fluid flow that do not result in particle acceleration to X-ray-emitting energies as a feasible explanation for the few moving, radio-only knots. \cite{Hardcastle2003} and \cite{Wykes2015} proposed stars of NGC~5128 with high mass-loss rates as possible obstacles in the jet. Detailed numerical tests of interactions between the jet and winds from stars embedded within it, for 3~Gyr old asymptotic giant branch (AGB) stars at their highest mass-loss rates, will be discussed in a forthcoming paper \citep{Wykes2018a}. 

In the present work, we use \chandra{} observations taken over 15~yr to investigate morphological evolution and proper motions of the A, B, and C knot complexes (Figure~\ref{fig:cena}), which are located in the inner 75\arcsec{} (1.38 kpc) of the jet (referred to as the `inner jet' in what follows). The remainder of the paper is structured as follows. In Section~\ref{sect:observation}, we provide the details of the \chandra{} observations and data reduction. Section~\ref{sect:diffmap} is devoted to the construction of various maps, with an emphasis on difference maps and statistical uncertainties. Proper motion measurements are pursued in Section~\ref{sect:motion}. Some physical consequences of our results are discussed in Section~\ref{sect:discuss}: results for the proper motion are compared with other FR~I systems, and brightness changes of the knots are related to the magnetic field strengths and expansion rates to investigate the primary mechanism for fading. Lastly, the results are compared with predictions from various knot formation models.

%%%%%%%%%%%%%%%%%
\section{Data Acquisition and Reduction} 
\label{sect:observation}

Prior analysis of X-ray emission from Cen\,A has shown it to be dominated by relatively low-temperature thermal and synchrotron emission that collectively peak at spectral energies below 1.0 keV \citep[and references therein]{Goodger2010}. To optimize the count rates, we used \chandra{} observations taken with the S3 chip of the Advanced CCD Imaging Spectrometer (ACIS), as it provides the greatest soft X-ray spectral sensitivity available with the instrument. Utilizing the same instrument for all observations also reduces the systematic uncertainty of any derived quantities from the image set. 

\begin{table}
	\caption{\textit{Chandra} Observations of Centaurus\,A Used}
	\label{table:obs}
	\begin{tightcenter}
		\begin{tabular}{ c c c c }
		\hline
		\hline
		ObsID & Instrument & Date & $t_{\rm exp}^{a}$ (ks)\\
		\hline
			02978 & ACIS-S & 2002 Sep 3 & 44.6 \\
			03965 & ACIS-S & 2003 Sep 14 & 48.9 \\
			10722 & ACIS-S & 2009 Sep 8 & 49.5 \\ 
			19521 & ACIS-S & 2017 Sep 17 & 14.8 \\
			20794 & ACIS-S & 2017 Sep 19 & 106.8 \\
		\hline
	\end{tabular}
	\end{tightcenter}
	{${}^{a}$ Net exposure after background flare removal}
\end{table}

Since any proper motion within the jet of Cen\,A will be small relative to \chandra{}'s resolution, the inner jet must be at the best focus possible to minimize blurring due to the point spread function (PSF). Cen\,A was observed with \chandra{} on 2002 September 03 with the inner jet centered on the S3 chip of ACIS in FAINT mode. Subsequent \chandra{} observations that used the same telescope configuration and aimpoint were performed in 2003 and 2009. Observations were also taken in 2017 using the same configuration for the express purpose of studying proper motions against the archival data. Our analysis relies mainly on the 2002/2003 and 2017 observations in order to investigate variability and proper motion over the largest available timespan, using the data with the best available spatial resolution. The 2009 observation was used to better sample variability detected from bright sources (Section~\ref{sect:diffmap}). In a companion paper \citep{Wykes2018b}, these observations are used to determine the pressure profile of the galactic atmosphere in the vicinity of the jet.

A complete list of the observations utilized in the analysis is provided in Table~\ref{table:obs}. All data were reprocessed using \ciao{} 4.9 with \caldb{} 4.7.6 \citep{Fruscione2006}. The \ciao{} task {\tt deflare} was used with default settings to remove background flares from the observations. The resulting cleaned exposure times are also shown in Table~\ref{table:obs}. Readout streaks in the images caused by the bright nucleus were removed with {\tt acisreadcorr}. The {\tt readout\_bkg} routine was employed to estimate the distribution of `out-of-time' events for each observation. 

Contaminant build-up over time on the ACIS optical blocking filter has caused a significant reduction in throughput at photon energies below $\sim$\,0.9 keV\footnote{See Section 6.5 of the ``Proposers Observatory Guide", \\ \url{http://cxc.harvard.edu/proposer/POG/html/chap6.html\#tth_sEc6.5. }}. To minimize the impact of this on the response, all images were binned with a lower-energy bound of 0.9 keV. An upper-energy bound of 2.0 keV was used to avoid PSF broadening that occurs at higher energies. The PSF for the selected energy range is well-characterized and stable over time\footnote{See Section 6.6 of the ``Proposers' Observatory Guide", \\ \url{http://cxc.harvard.edu/proposer/POG/html/chap6.html\#tth_sEc6.6}}. The average count rate for the 0.9-2.0 keV band of the 2017 data set was shown to be equivalent to the 2002/2003 count rate after correcting for minor differences in \chandra's sensitivity. 

\begin{figure}
	\begin{tightcenter}
	\includegraphics[width=0.46\textwidth]{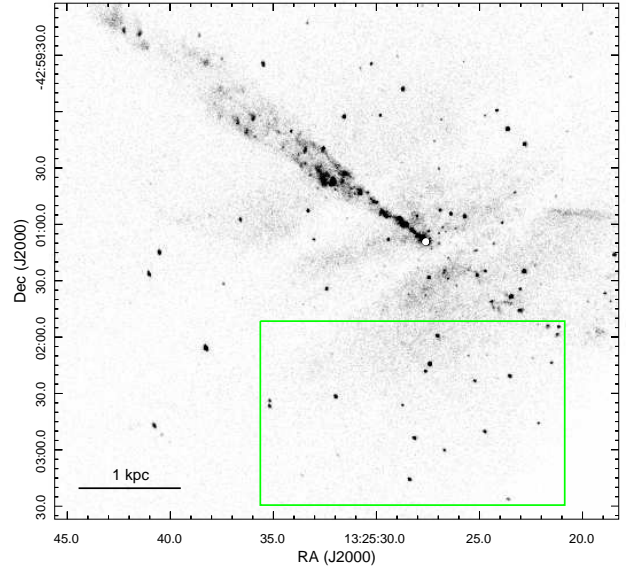}
	\end{tightcenter}
    \caption{0.9--2.0 keV \chandra{} image of the Centaurus\,A jet and 
    		its surroundings. ACIS-S observations listed in 
		Table~\ref{table:obs} were co-added for the image. Pixel size is 
		0.492\arcsec, and the image has been smoothed with an 8 pixel 
		RMS Gaussian filter. Overlaid is a rectangular region (green) used for the 
		cross-correlation analysis (Section~\ref{sect:observation}). The region 
		was selected to be close to the detector aimpoint. Point 
		sources in the region were used to correct for differences in astrometry 
		between the various observations. The AGN has been masked in the
		image.}
	\label{fig:region}
\end{figure}

\begin{table*}
	\caption{Average Change in X-Ray Brightness, On-source Background Subtraction}
	\label{table:diffbright}
	\begin{tightcenter}
		\begin{tabular}{ c c c c c c }
		\hline
		\hline
		Knot & Integrated S/N & \multicolumn{3}{c}{Brightness 
			(10$^{-6}$ photons cm$^{-2}$ s$^{-1}$)} & Average Change \\
		\cline{3-5}
		& [$(S/N)^2/\rm pixels$] & 2002/2003 Epoch & 2009 Epoch & 2017 Epoch 
			& In Brightness$^{a,b}$ \\
		\hline
		AX1A & 187.1/161 & $10.2 \pm 0.4$ & $12.2 \pm 0.7$ & $13.0 \pm 0.5$ 
			&  $+27 \pm 10\%$ \\
		AX1C & 168.8/141 & $17.5 \pm 0.6$ & $15.8 \pm 0.8$ & $13.6 \pm 0.5$ 
			&  $-23 \pm 6\%$ \\ 
		AX2 &108.4/123 & $2.7 \pm 0.2$ & $2.9 \pm 0.3$ & $2.2 \pm 0.2$ 
			&  $-20 \pm 17\%$ \\
		AX3 & 119.4/140 & $2.8 \pm 0.2$ & $2.5 \pm 0.3$ & $2.2 \pm 0.2$ 
			& $-22 \pm 17\%$ \\
		AX4 & 183.7/264 & $3.0 \pm 0.2$ & $3.5 \pm 0.4$ & $2.9 \pm 0.2$ 
			& $-6 \pm 19\%$ \\ 
		AX5 &120.0/267 & $1.8 \pm 0.2$ & $2.3 \pm 0.3$ & $1.7 \pm 0.2$ 
			& $-14 \pm 26\%$ \\
		AX6 & 197.5/234 & $4.5 \pm 0.3$ & $4.2 \pm 0.4$ & $5.7 \pm 0.3$ 
			& $+27 \pm 15\%$ \\ 			
		BX1 & 161.4/271 & $2.2 \pm 0.2$ & $2.1 \pm 0.3$ & $2.3 \pm 0.2$ 
			& $+4\pm 24\%$ \\
		BX2A & 78.5/192 & $1.0 \pm 0.1$ & $1.4 \pm 0.2$ & $1.0 \pm 0.1$ 
			& $-9 \pm 38\%$ \\
		BX2 & 281.0/267 & $15.2 \pm 0.5$ & $14.1 \pm 0.7$ & $13.0 \pm 0.5$ 
			& $-15 \pm 7\%$ \\
		BX4 & 149.7/222 & $2.3 \pm 0.2$ & $2.7 \pm 0.3$ & $2.6 \pm 0.2$ 
			& $+15 \pm 23\%$ \\
		BX5 & 363.1/778 & $4.9 \pm 0.3$ & $4.9 \pm 0.4$ & $4.6 \pm 0.3$ 
			& $-14 \pm 18\%$ \\
		CX1 & 256.3/800 & $2.7 \pm 0.2$ & $3.6 \pm 0.4$ & $3.0 \pm 0.2$ 
			& $+8 \pm 33\%$ \\
		CX2 & 72.9/78 & $1.7 \pm 0.2$ & $2.2 \pm 0.3$ & $2.1 \pm 0.2$ 
			& $+21 \pm 24\%$ \\
		\hline
	\end{tabular}
	\end{tightcenter}
	{${}^{a}$Defined as ($B_{2017}$ - $B_{2002/2003}$)/ $B_{2002/2003}$, where is $B$ is brightness \\
	${}^{b}$ Positive values indicate an increase in brightness toward the current epoch; negative values indicate a decrease in brightness}
\end{table*}

Investigation of proper motion requires the images of Cen\,A to be co-aligned to high accuracy. Cross-correlation was used to determine and correct any residual astrometric offsets between the observations \citep{Snios2018b}. A rectangular region of $165\arcsec \times 100\arcsec$ centered on the field of point sources to the south of Cen\,A was then defined (see Figure~\ref{fig:region}). This region was selected to be close to the detector aimpoint in order to minimize the PSF and provide the highest possible accuracy for the image co-alignment. No features intrinsic to the non-thermal jet/counterjet emission of Cen A were included in the region to avoid alignment biasing from potential temporal variations. In total, 24 unique point sources that were present in all observations were detected within the selected region. This provided adequate statistics to ensure accurate alignment of the data sets. The region was also varied by size, position, and orientation numerous times to ensure the resulting offsets were not biased by the region selection. All offsets reported in this work were found to be insensitive to these variations.

An image of the selected region was generated for each ObsID. For this method, ObsID 20794 was selected as a reference image for its high total exposure. A two-dimensional cross-correlation function between ObsID 20794 and each remaining image was generated. A least-squares fit of a two-dimensional Lorentzian was applied to the cross-correlation function to determine the relative offset of each image to the reference. The astrometric shift needed to correct this offset was applied to the data using the \ciao{} {\tt wcs\_update} routine, requiring a root mean square translation of ($\Delta {x}_{\rm rms}$, $\Delta{y}_{\rm rms}$) = (0.32\arcsec, 0.64\arcsec). Here, the directions of the angular offsets $(\Delta x, \Delta y)$ correspond to the directions of R.A. and decl., respectively. The goodness of the alignment was tested by comparing positions of the surrounding point sources using the centroid positions from the {\tt dmstat} routine; pairs were found to agree within $0.01\arcsec$. The agreement is significantly more precise than the accuracy of the estimated proper motion (see Section~\ref{sect:motion}).

Generation of accurate exposure map corrections for Cen\,A is difficult given the significant spectral variations over the system due to the presence of the dust lane near the central AGN. To avoid introducing uncertainties into the final images owing to assumptions of the spectral model, multiple exposure maps were created in 0.1 keV slices using the average effective area of each slice. Exposure maps were generated using a sub-pixel binning of $0.123\arcsec \rm\ pix^{-1}$ with the {\tt mkexpmap} \ciao{} command. The exposure-corrected image slices were then co-added to produce a final, exposure-corrected image for each observation. The final images were binned over a \mbox{0.9--2.0} keV energy band, and the exposure-corrected images are in units of $\rm photon\ cm^{-2}\ s^{-1}$. A merged, exposure-corrected image of Cen\,A, with the knots labeled, is shown in Figure~\ref{fig:cena}. Following \cite{Hardcastle2007}, we define a knot as any compact feature in the jet that is distinguished (by a factor $\ga$ 2) in surface brightness from its surroundings, and has a radius $< 2$\arcsec.

%%%%%%%%%%%%%%%%%%%%%%%%%%%
\section{Difference Maps}
\label{sect:diffmap}

 \begin{figure*}
	\begin{tightcenter}
	\includegraphics[width=0.50\textwidth]{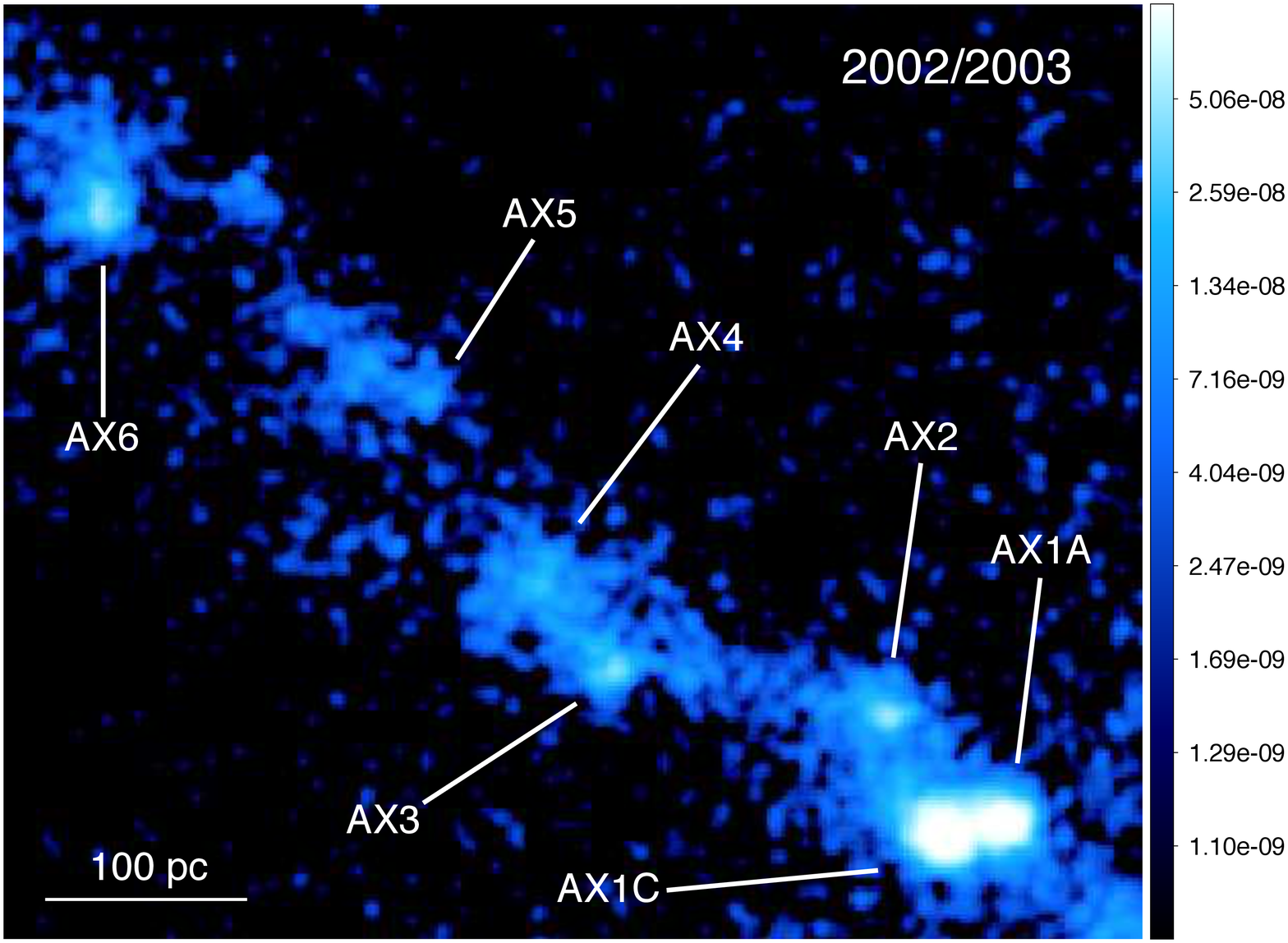}\includegraphics[width=0.50\textwidth]{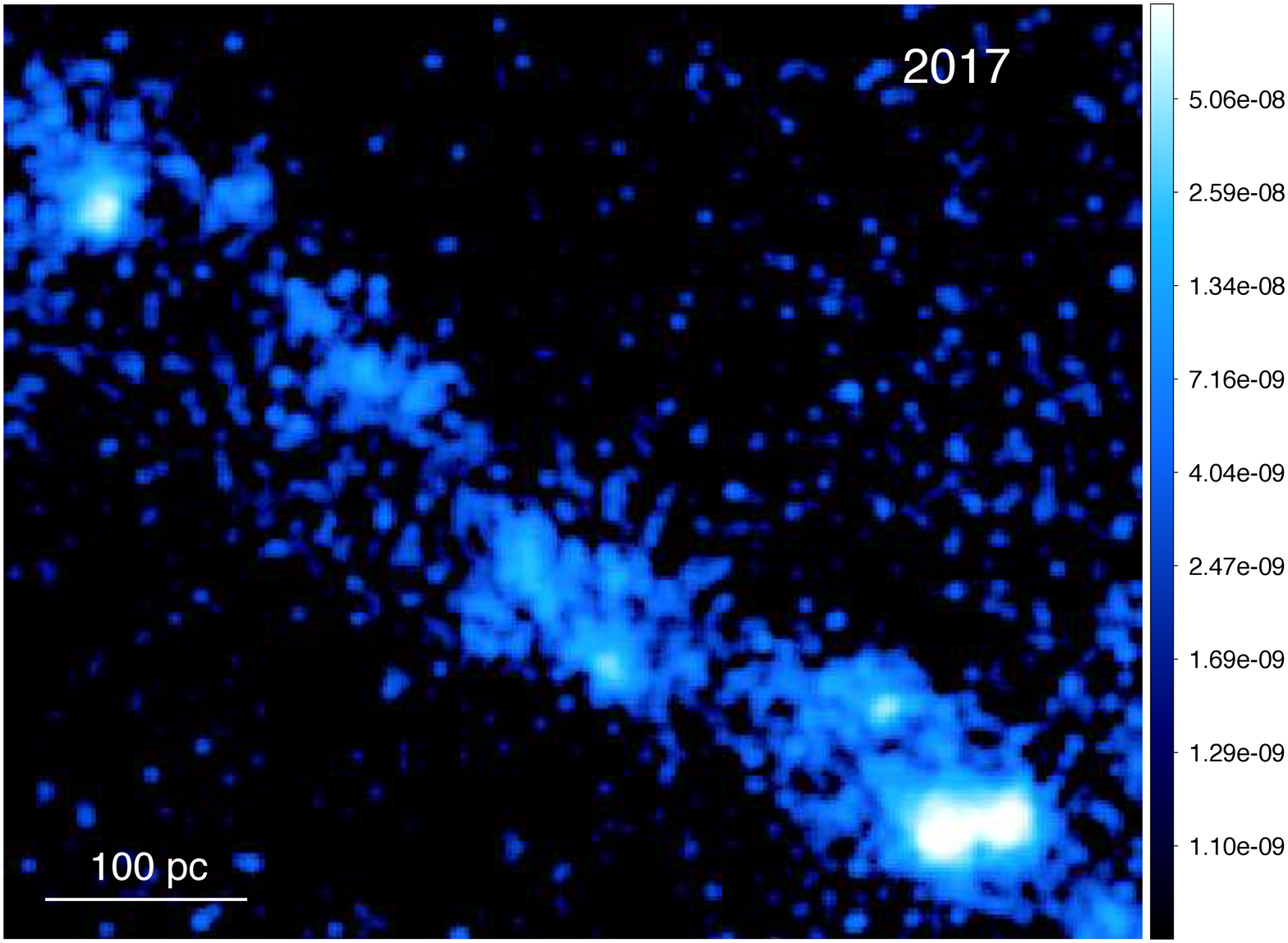} \\
	\includegraphics[width=0.50\textwidth]{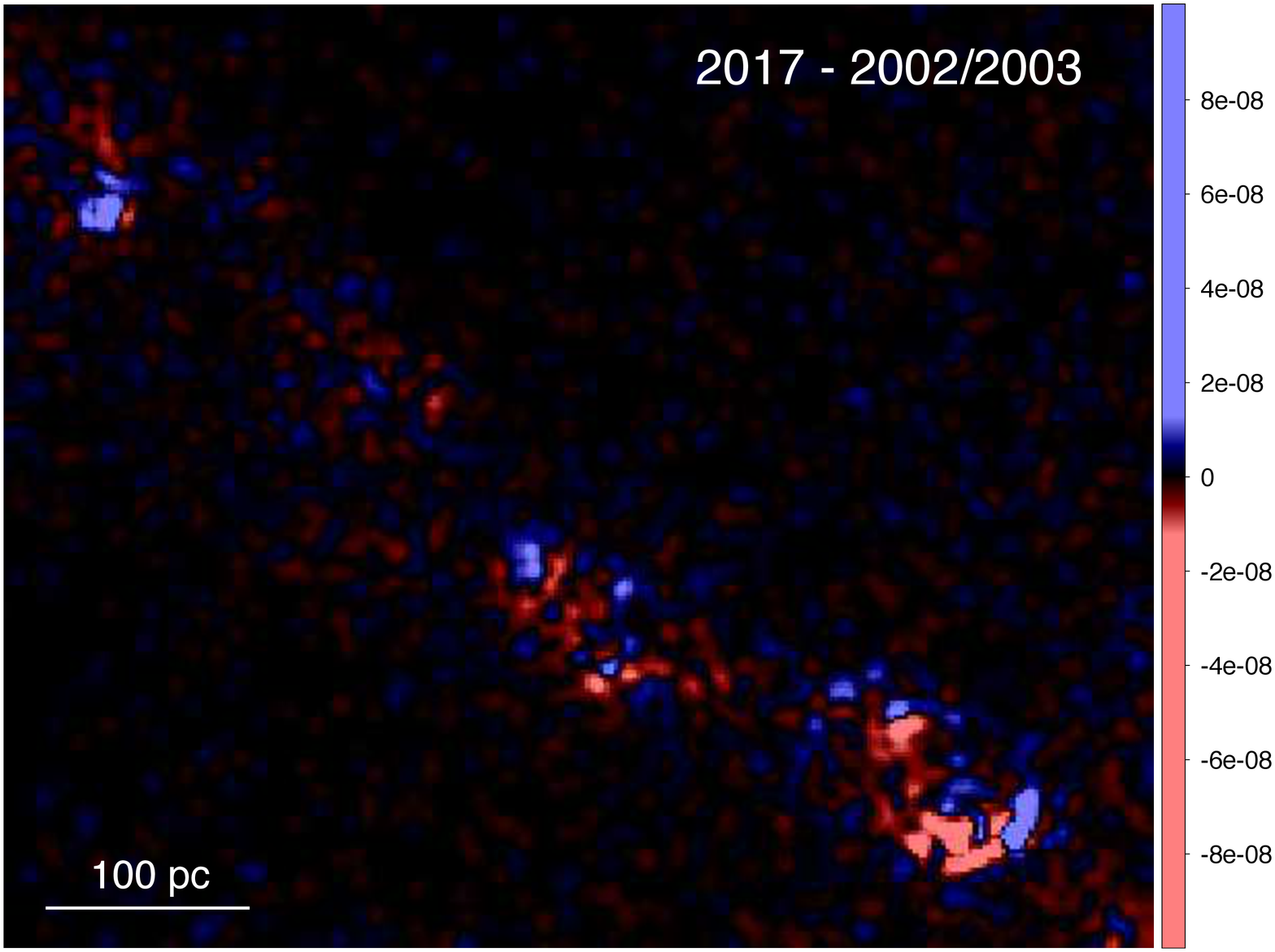}\includegraphics[width=0.50\textwidth]{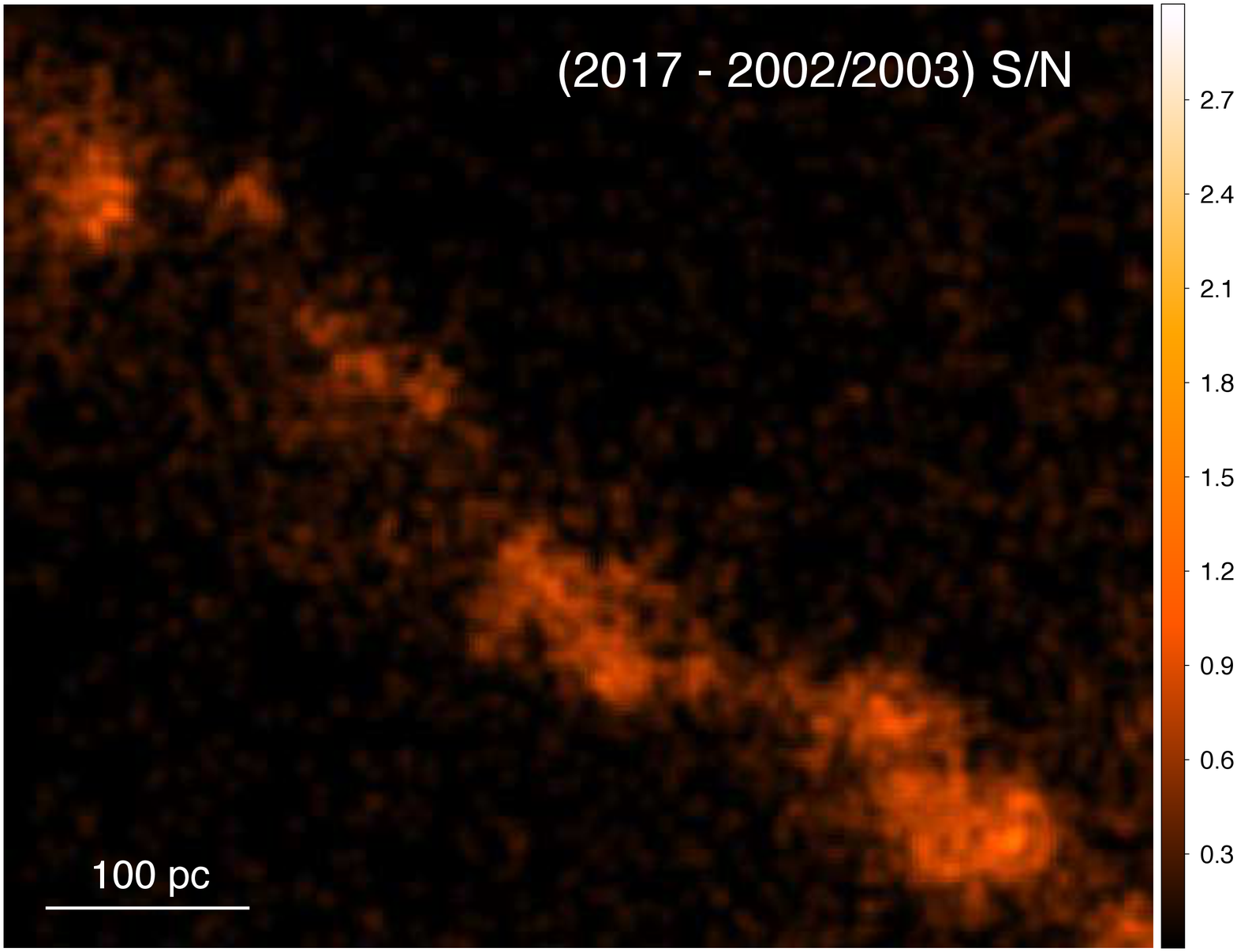} 
	\end{tightcenter} 		
    \caption{Images at 0.9--2.0 keV (upper panels), difference map 
    		(lower left panel), and S/N map (lower right panel) are shown of the Knot A
    		complex in the Centaurus\,A jet. The images are binned on a scale of 
		$0.123\arcsec \rm\ pix^{-1}$ and smoothed with a 3 pixel RMS 
		Gaussian. The exposure-corrected images are in units 
		of $\rm photon\ cm^{-2}\ s^{-1}$. In the difference map, the red regions 
		are areas that are brighter in the 2002/2003 data set, while the 
		blue regions are brighter in the 2017 data.}
	\label{fig:diffa}
\end{figure*}

 \begin{figure*}
	\begin{tightcenter}
	\includegraphics[width=0.50\textwidth]{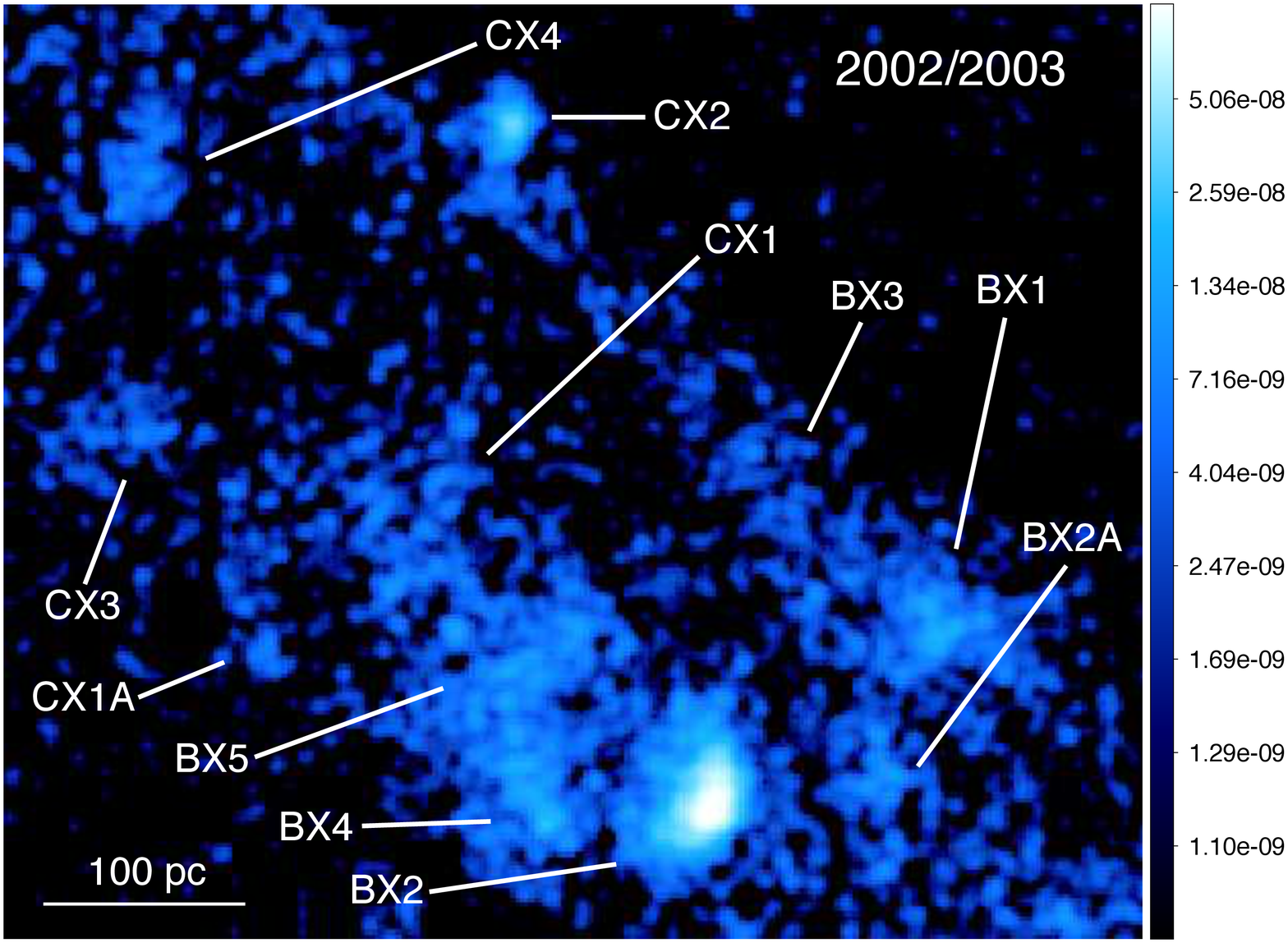}\includegraphics[width=0.50\textwidth]{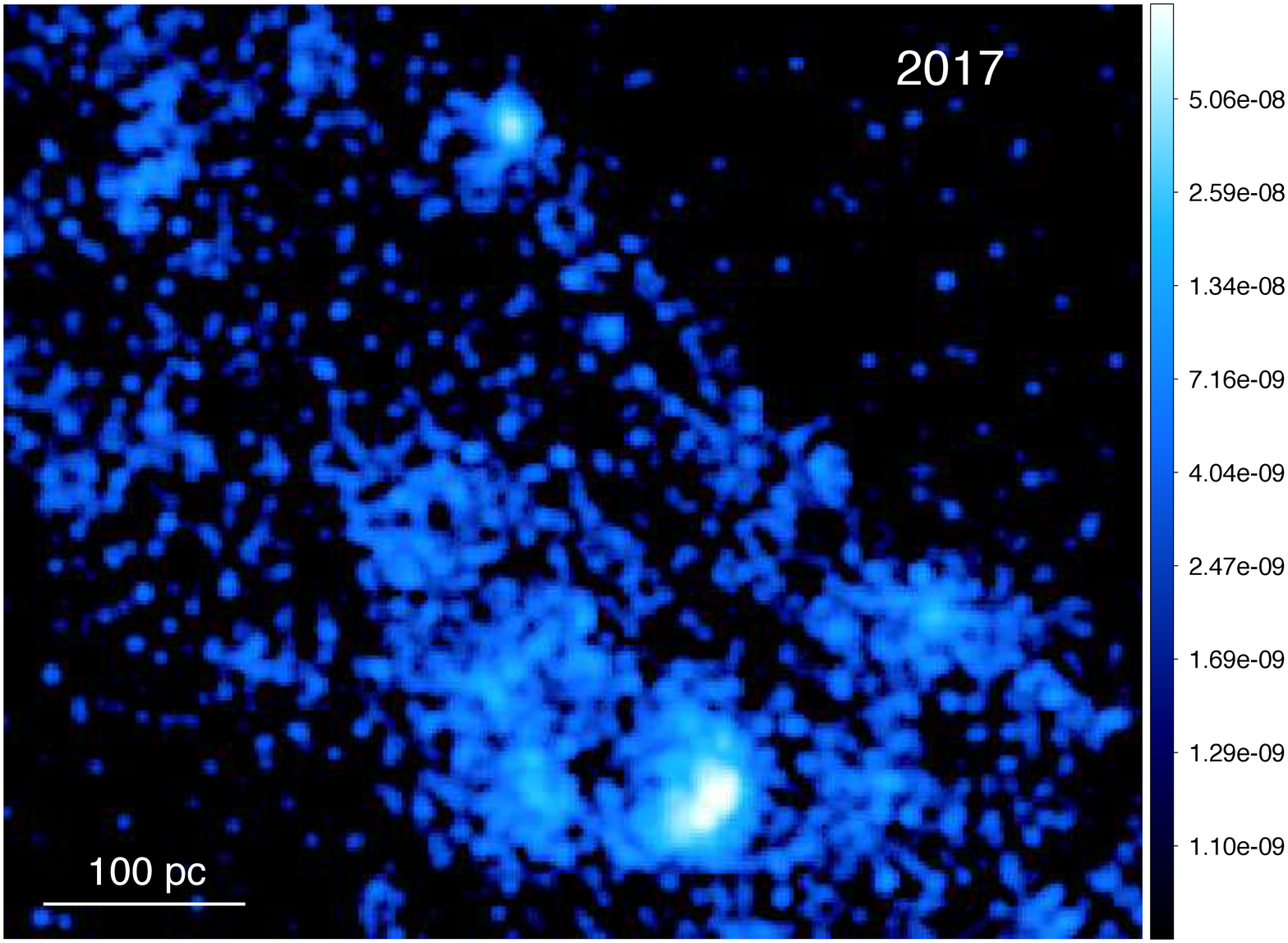} \\
	\includegraphics[width=0.50\textwidth]{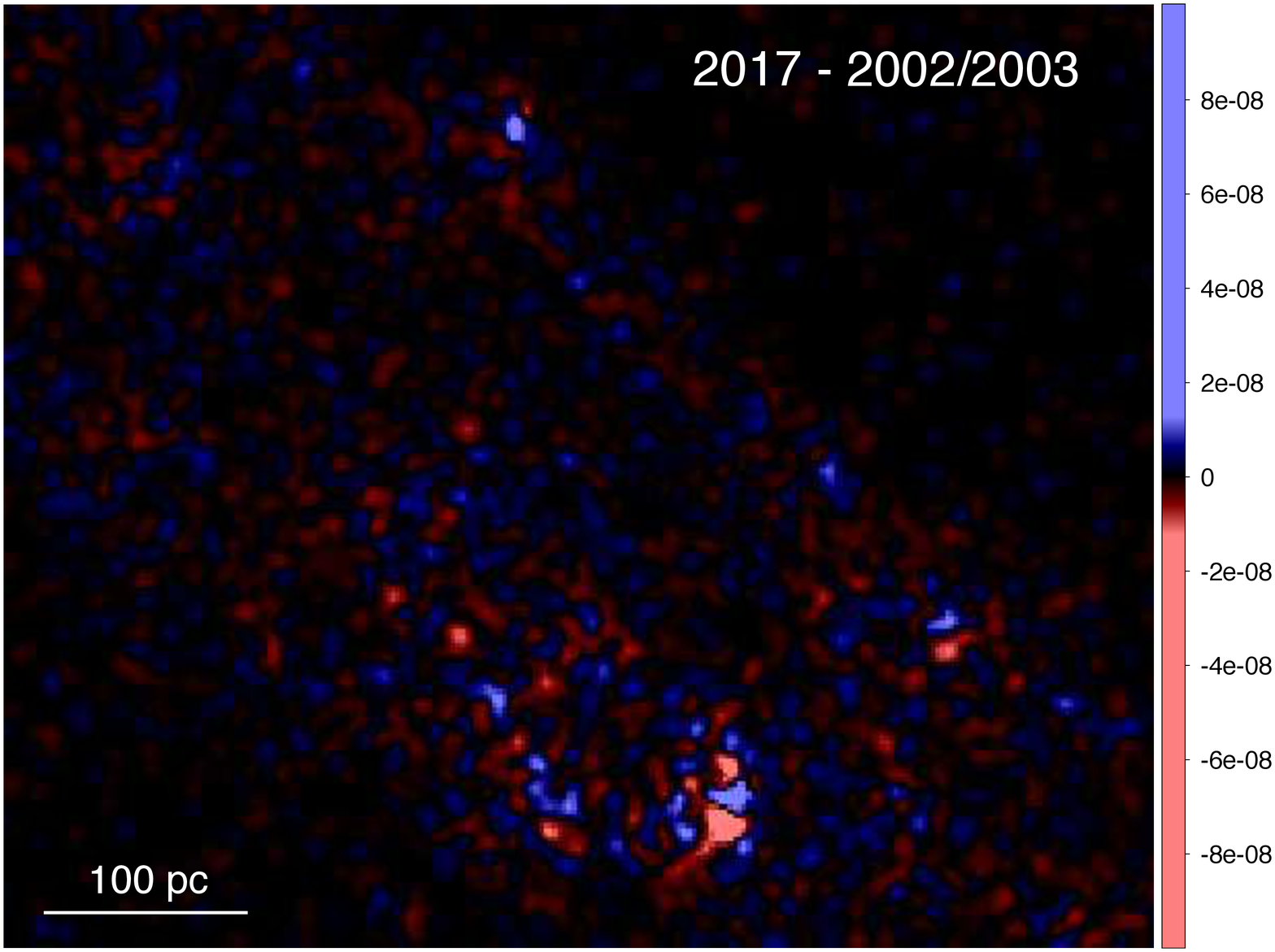}\includegraphics[width=0.50\textwidth]{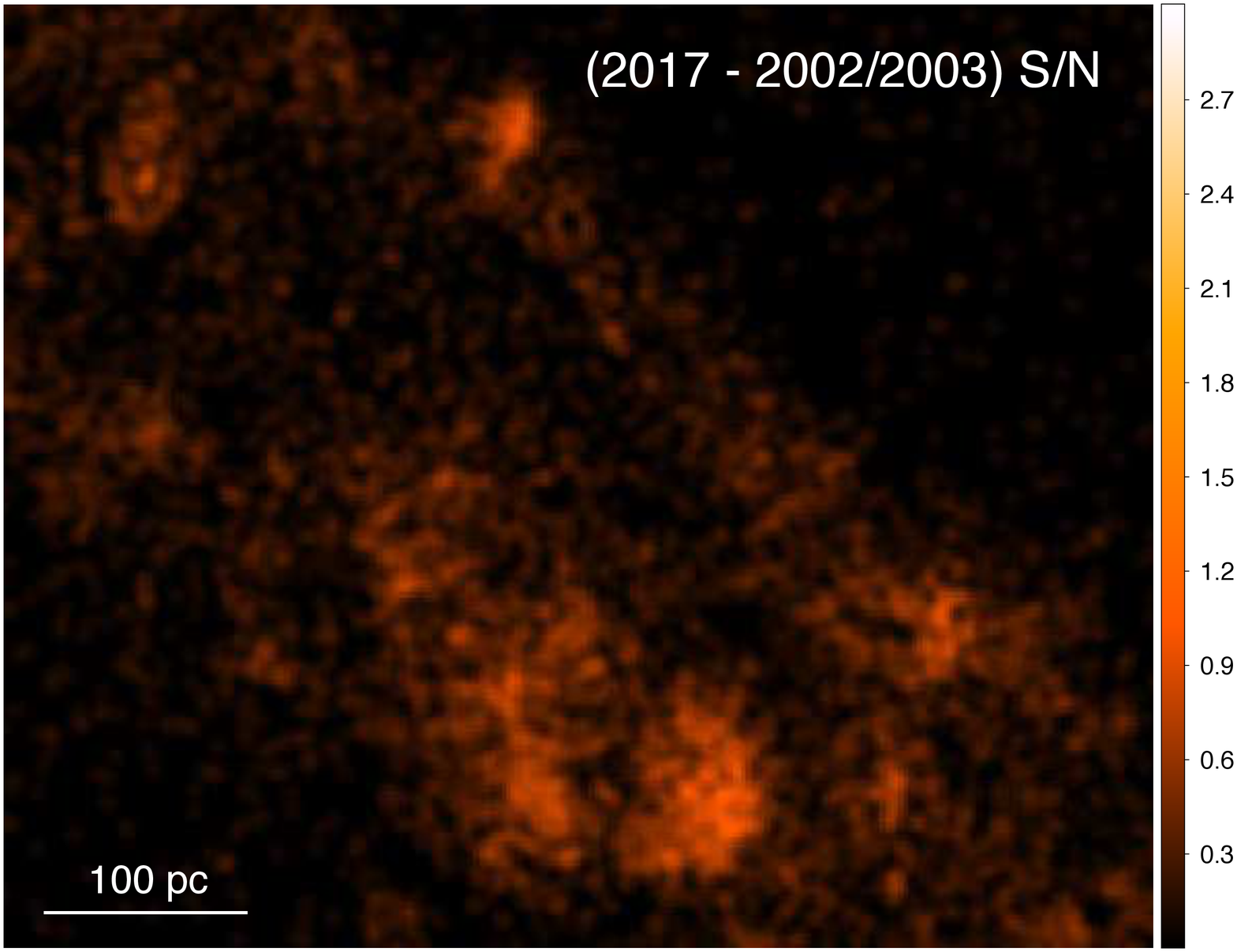} 
	\end{tightcenter} 		
    \caption{Images at 0.9--2.0 keV (upper panels), difference map 
    		(lower left panel), and S/N map (lower right panel) are shown of the Knot B
    		and Knot C complexes in the Centaurus\,A jet. The images are binned on 
		a scale of $0.123\arcsec \rm\ pix^{-1}$ and smoothed with a 3 
		pixel RMS Gaussian. The exposure-corrected images are in units 
		of $\rm photon\ cm^{-2}\ s^{-1}$. In the difference map, the red regions 
		are areas that are brighter in the 2002/2003 data set, while the 
		blue regions are brighter in the 2017 data.}
	\label{fig:diffb}
\end{figure*}

The reprocessed images described in Section~\ref{sect:observation} were separated into 2002/2003 and 2017 epochs to probe the longest available timescale for variability and proper motion. ObsIDs 02978 and 03965 were merged with the \ciao{} routine {\tt merge\_obs} to create the 2002/2003 data set, while ObsIDs 19521 and 20794 were merged for the 2017 data set. A difference map of the two epochs was generated using {\tt dmimgcalc}. The final images and difference maps are presented in Figures~\ref{fig:diffa} and \ref{fig:diffb}.

Inspection of those point sources adjacent to the jet of Cen\,A that remained constant in brightness over time reveals no notable morphological features in the difference maps, reaffirming that the astrometric and exposure corrections are accurate. Variations in brightness and morphology were observed throughout the Knot A, B, and C complexes. Variations in knots located further downstream and the counterjet knots were found to be consistent with the observed background fluctuations and are consequently not shown.

The uncertainty in the difference map must be quantified to determine the significance of any changes in the jet. If the raw counts for a pixel in the two 0.9--2.0 keV exposures are $N_1$ and $N_2$ and the corresponding exposure corrections are $c_1$ and $c_2$, then the value in
the difference image is $c_2 N_2 - c_1 N_1$. We estimate the signal-to-noise ratio (S/N) as
\begin{equation} 
{\rm S/N} = \frac{|c_2 N_2 - c_1 N_1|}{\sqrt{c_1^2 N_1 + c_2^2 N_2}}\,. 
\end{equation} 
S/N maps computed from this method are shown in Figures~\ref{fig:diffa} and \ref{fig:diffb}. Regions were defined surrounding each knot based on the knot definition criteria from Section~\ref{sect:observation}, and integrated S/Ns were subsequently calculated for each of these regions. The integrated $\chi^2$ for a region is defined as 
\begin{equation} 
\sum\limits_{i}^{N_{P}} ({\rm S/N})_{i}^2\,, 
\end{equation} 
where ${\rm(S/N)}_{i}$ is the S/N of a pixel and $N_{P}$ refers to the total number of pixels within the region. Values of $\chi^2$ and $N_{P}$ for each region are shown in  Table~\ref{table:diffbright}. In the absence of a signal, these values would have a $\chi^2$ distribution with $N_{P}$ degrees of freedom.  Knots AX1A, AX1C, and BX2 each have $\chi^2$ values that are inconsistent with zero difference above the $3 \sigma$ level. The remaining knots all lie below the $3\sigma$ threshold, while variations throughout the diffuse emission of the jet lie below the $1\sigma$ threshold. 

Based on \chandra{} ACIS-S calibrations, the primary source of systematic error in the difference map is due to the growing amount of contaminant on the ACIS optical blocking filter. This contaminant layer is known to contribute a maximum systematic count rate error $< 5\%$\footnote{See ``ACIS QE Contamination", \\ \url{http://cxc.harvard.edu/ciao/why/acisqecontamN0010.html}}. To determine which observed variations exceeded this uncertainty threshold, percentage changes in brightness were estimated for each area of significant brightness change. Regions were again defined surrounding each knot, and the brightness was estimated for each knot in the 2002/2003, 2009, and 2017 epochs. On-source annular regions were used for background-subtraction to account for the diffuse jet emission.  Brightness uncertainties were estimated from Poisson noise. An average brightness change was calculated by taking the difference of the 2017 and 2002/2003 epochs and then dividing the result with the 2002/2003 brightness. This method provides information on the change in brightness relative to the initial epoch. The brightness and average brightness change for each examined knot are provided in Table~\ref{table:diffbright}. 

Several features in the difference map show significant variations in brightness. Clear variations are observed in the Knot A complex, as shown in Figure~\ref{fig:diffa}, with knots AX1A and AX1C demonstrating the most significant brightness and morphological variations. Structures surrounding the AX1 knots also appear to vary in brightness, though this region was found to possess a S/N~$<$~2 in the difference map after accounting for the variability from the knots. Knots AX2-AX6 appear to exhibit brightness changes, but the low integrated S/N does not allow for a robust quantification. Moving down the jet, the Knot B and Knot C complexes also display variations in the difference images (see Figure~\ref{fig:diffb}), albeit at a reduced S/N when compared with the Knot A complex. Knot BX2, which has the second-highest flux density of all the observed knots, shows significant brightness and morphological shifts, while there are no other features in the B and C complexes with sufficient S/N to conclude that they have varied. 

%%%%%%%%%%%%%%%%%%%%%%%%%%%
\section{Proper Motion}
\label{sect:motion}

\begin{figure}
	\begin{tightcenter}
	\includegraphics[width=0.47\textwidth]{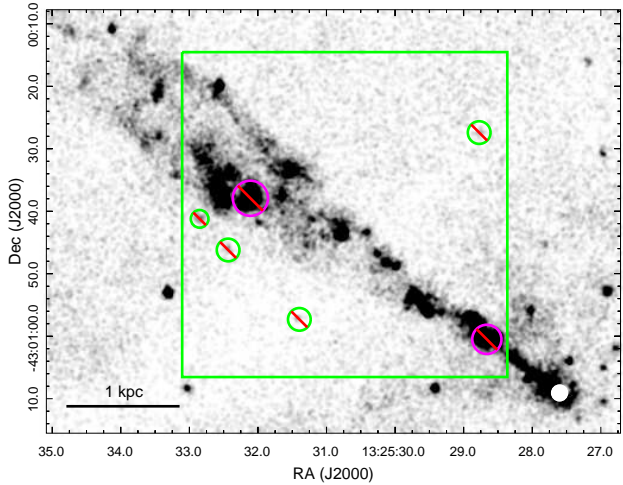}	
	\end{tightcenter} 		
    \caption{0.9--2.0 keV \chandra{} image of the Centaurus\,A inner jet. 
    		The ACIS-S observations listed in Table~\ref{table:obs} were 
		co-added for the image. Pixel size is 0.492\arcsec, and the 
		image has been smoothed with an 8 pixel RMS Gaussian filter.
		Overlaid is a $50\arcsec \times 50 \arcsec$ region (green square)
		used for the cross-correlation analysis of the jet (Section~\ref{sect:motion}). 
		Adjacent point sources were masked for both the first and second 
		cross-correlation fits (green circles), while a second fit also has 
		knots AX1A, AX1C, and BX2 masked (magenta circles).}
	\label{fig:mask}
\end{figure}

 \begin{figure*}
	\begin{tightcenter}
	\includegraphics[width=0.50\textwidth]{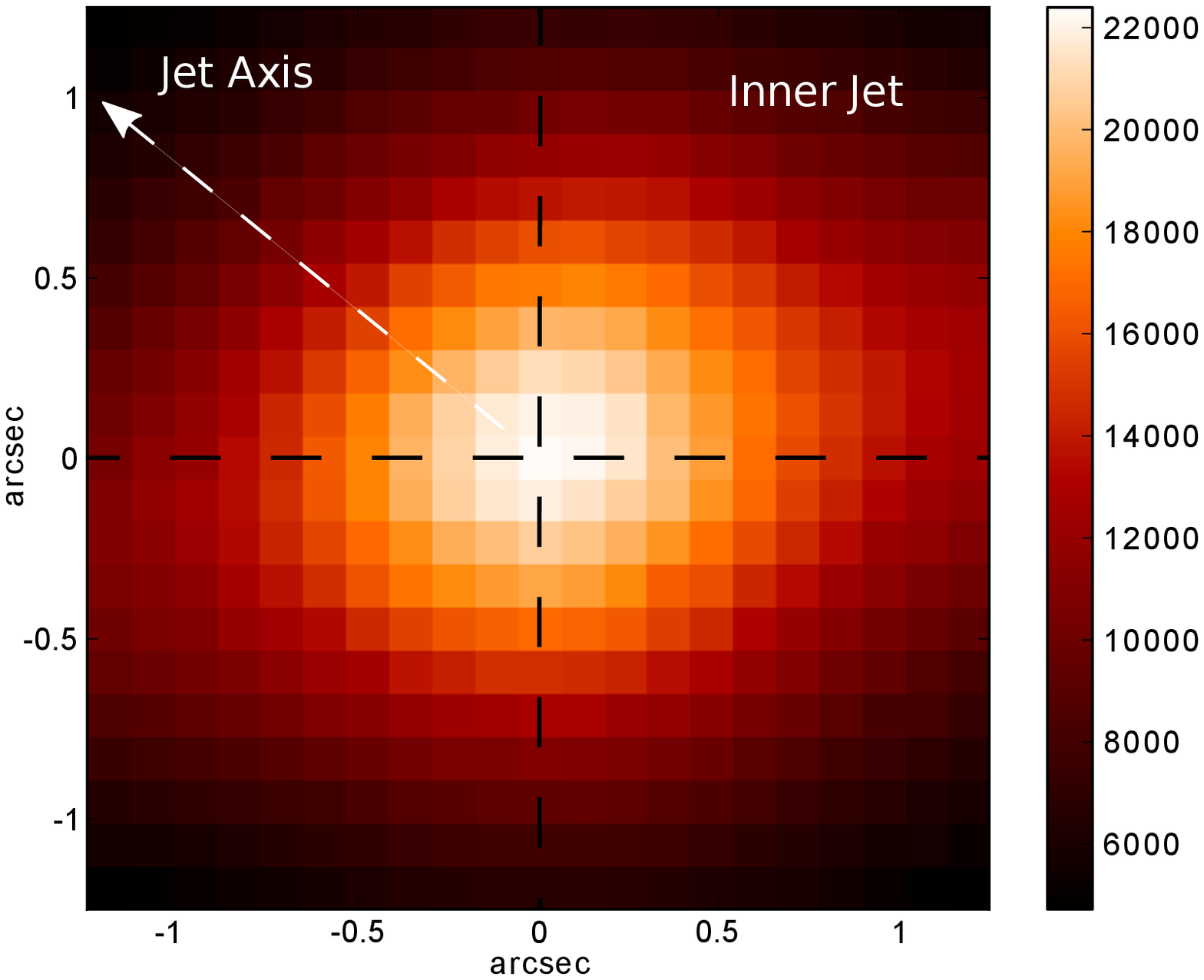}\includegraphics[width=0.50\textwidth]{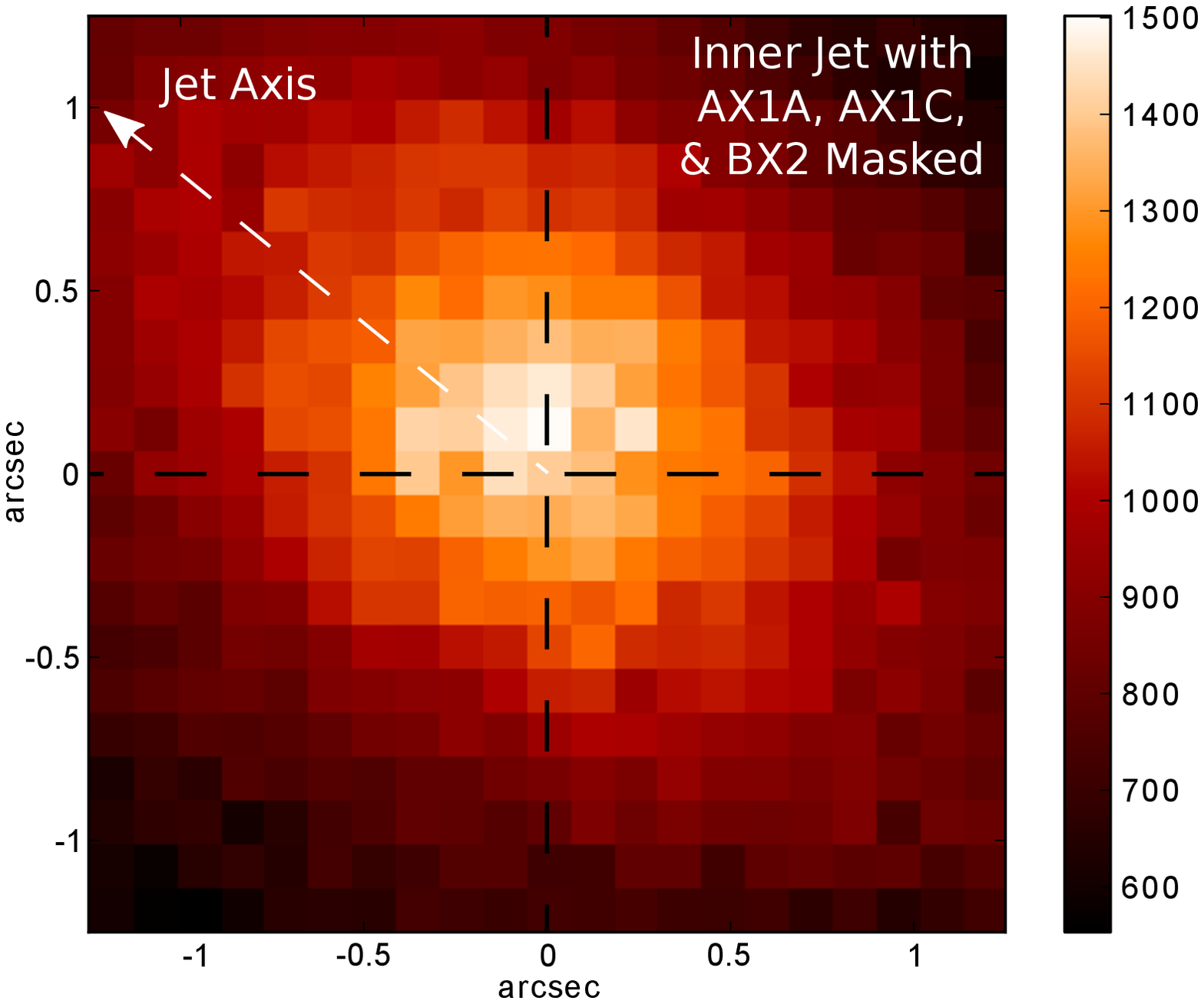}
	\end{tightcenter} 		
    \caption{Plots of the cross-correlation functions from Section~\ref{sect:motion}, 
    		zoomed in on the absolute maxima.
    		The pixel size corresponds to $0.123\arcsec \rm\ pix^{-1}$. 
    		The first fit masked adjacent point sources to the jet (left), while the 
		second fit also masked knots AX1A, AX1C, and BX2 (right). 
		Masking the three bright knots reveals a shift along the jet axis}
	\label{fig:knot_cc}
\end{figure*}

\begin{figure*}
	\begin{tightcenter}
	\includegraphics[width=0.95\textwidth]{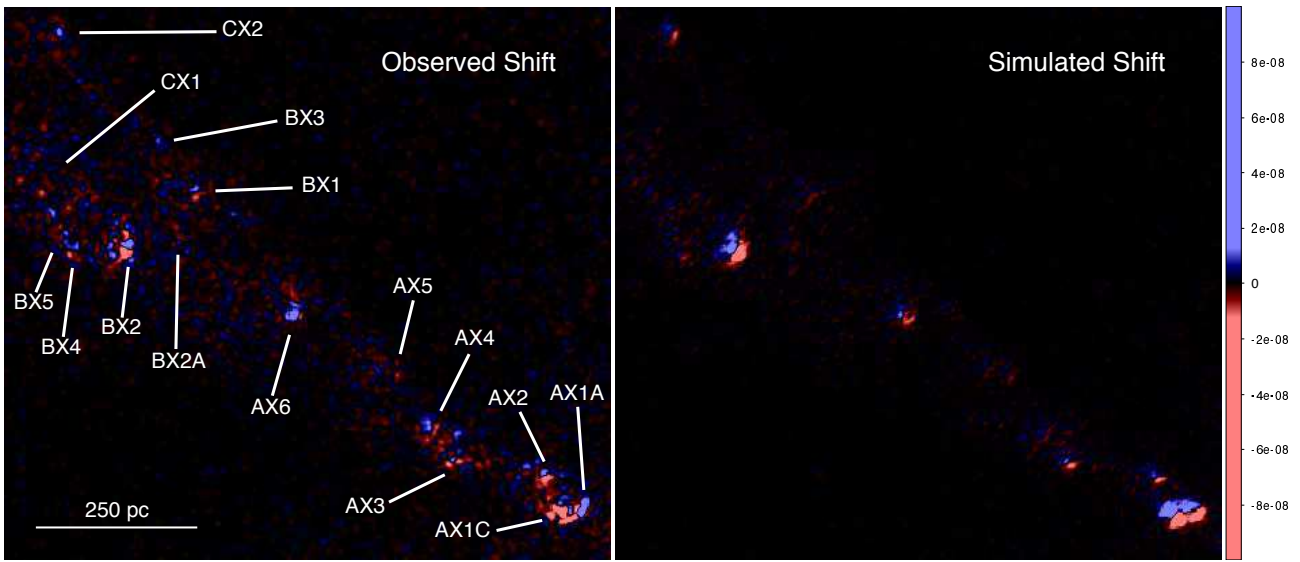}	
	\end{tightcenter} 		
    \caption{Comparison between the observed difference map 
    		(left) and a simulated difference map assuming a proper motion 
		of 0.68c along the jet axis (right). The images are binned on 
		a scale of $0.123\arcsec \rm\ pix^{-1}$, smoothed with a 3 
		pixel RMS Gaussian, and are in units of 
		$\rm photon\ cm^{-2}\ s^{-1}$. The red regions correspond 
		to brighter areas in the 2002/2003 data set, while the blue 
		regions correspond to brighter areas in the 2017 data set.}
	\label{fig:synth}
\end{figure*}

Here, we discuss evidence for motion of X-ray knots in the Cen\,A jet. An object at the distance of Cen\,A moving at a projected speed $c$ would have a proper motion of $\simeq 0.25\arcsec$ over a 15 year timespan. Thus, we should expect the resultant changes in the X-ray images to be subtle. In the difference maps (Figures~\ref{fig:diffa} and \ref{fig:diffb}), a knot that has moved outward significantly should show positive residuals at its outer margin and negative residuals at its inner margin. Inspection of the difference maps reveals that the brightest knots (AX1A, AX1C, and BX2) show no convincing evidence of downstream movement along the jet axis. As a further check, the position of each of these knots was determined using the \ciao{} {\tt dmstat} routine and the results for the two epochs compared. The RMS shift in position for the three knots was measured to be 0.03\arcsec, which is smaller than our estimate of the uncertainty (derived below). In contrast, potential evidence of motion is observed for several other knots, like AX2 or AX6, but their low count statistics make it impossible to directly measure the projected speed of any individual knot. 

Although none of the fainter knots can be located individually with sufficient precision to detect proper motions, there are indications that several of the fainter X-ray knots have moved downstream along the jet. To look for evidence of collective motion in the jet, we computed the cross-correlation function of images for the 2002/2003 and 2017 epochs. Analogous to the method used to co-align the images (Section~\ref{sect:observation}), we first defined a $50\arcsec \times 50\arcsec$ region of the inner jet, which is shown in Figure~\ref{fig:mask}. All point sources within the field of view, but outside the jet, were masked. Two different regions were selected for this analysis: a region using the entire inner jet, and the same region but with knots AX1A, AX1C, and BX2 masked. These three knots were masked because they appear stationary to high significance and were therefore removed to unveil motions in the remainder of the structure in the jet.  Similar to the method outlined in Section~\ref{sect:observation}, the image region was varied by size, centroid position, and orientation numerous times to ensure the resulting offsets were not biased by the region selection. To ensure a comprehensive analysis, all possible alignments of the two images were sampled in the cross-correlation function.

With the three brightest knots included, the absolute maximum of the resulting cross-correlation function was found to be 20 times greater in amplitude than the background noise. A zoomed-in region of the cross-correlation function absolute maximum is shown in
Figure~\ref{fig:knot_cc}, left panel. All remaining local maxima detected in the cross-correlation function had an amplitude $< 5\%$ of the absolute maximum and were found to be coincident with background fluctuations. In fitting the peak with a two-dimensional Lorentzian profile, the image offset was found to be $(\Delta x, \Delta y) = (0.026\arcsec, 0.033\arcsec)$, consistent with these knots being stationary (see error discussion below). The directions of the angular offsets $(\Delta x, \Delta y)$ correspond to the directions of R.A. and decl., respectively.

Repeating the cross-correlation analysis with the three brightest knots excluded shows an order of magnitude reduction in the cross-correlation absolute maximum amplitude (Figure~\ref{fig:knot_cc}, right panel). This difference is expected given that the amplitude is directly dependent on the total counts and alignment of the images, and the three brightest knots are tightly localized sources within the jet that contribute 58\% of the total knot emission. The three knots therefore dominated the previous cross-correlation fit, and removal of them will allow us to probe for proper motion of the fainter knots and diffuse X-ray emission in the jet.

By masking the three brightest knots, the absolute maximum of the second cross-correlation function was observed to be 6 times greater than the background. The remaining local maxima detected had an amplitude $< 20\%$ of the absolute maximum and were consistent with noise. The peak shown in Figure~\ref{fig:knot_cc} (right panel) is therefore the most statistically significant feature found from the cross-correlation function. Fitting the peak with a two-dimensional Lorentzian profile, an offset for the remainder of the inner jet was found to be $(\Delta x, \Delta y) = (-0.166\arcsec, 0.046\arcsec)$, for a total shift of $0.17\arcsec$. It is noteworthy that there is no peak at zero offset in the right panel of Figure~\ref{fig:knot_cc}, implying that the greater part of the remaining substructure is moving at comparable speeds. The direction of the measured offset is at $75^\circ$ east of north projected on the sky. For the error estimate below, this is slightly more than $1\sigma$ away from the direction of the jet, at $55^\circ$ east of north projected on the sky. We therefore found that the fainter knots, and possibly other faint substructure in the jet, have moved along the jet in the flow direction.

The primary source of statistical error inherent to the proper motion estimate is the uncertainty in the cross-correlation fit. To quantify this uncertainty would require an accurate model of the X-ray image of the jet of Cen\,A. However, any model will introduce a significant source of systematic uncertainty that is difficult to assess and is therefore not advised for use when analyzing complex systems \citep{Peterson1998}. We instead opted to study the uncertainty by splitting the exposure time for each of the three available data sets (2002/2003, 2009, 2017) in half and cross-correlating each data set half with its complementary half. Since the complementary halves are essentially coeval, the observed shifts should be zero, apart from the statistical uncertainty of this method. While this approach only provides three test samples and datasets of half the length used for the actual measurement, the statistical properties of the test samples are mostly well-matched to the actual measurement and the method avoids the risk of introducing new systematic uncertainties. From this analysis, the RMS shift of the three data sets was 0.05\arcsec, with each data set randomly distributed about the zero-point. As this uncertainty analysis halves the total available exposure, it increases the statistical uncertainty and gives a value higher than the uncertainty from the proper motion calculation. The value of 0.05\arcsec{} should therefore be viewed as an upper limit on the statistical uncertainty for the proper motion estimate. 

An additional source of uncertainty in the result is from the 1 year spread in observations used for the 2002/2003 data set. To estimate this error, the 2002/2003 data set was first split into its two, separate observations. Each observation was then cross-correlated with the 2017 data set using the same regions from which the proper motion was estimated. A resulting 5\% spread in the previously reported proper motion shift was observed, which is small compared to the $\sim$\,30\% uncertainty from the cross-correlation fit. We therefore considered this systematic uncertainty to be negligible relative to other sources of error. 

The validity of the proper motion estimate was further examined by generating a synthetic difference map to compare with the observed map. A simulated observation was created by shifting the 2002/2003 image by 0.17\arcsec{} along the jet. A difference map of the simulated and actual observations was generated and then compared to the observed difference map (Figure~\ref{fig:synth}). Direct estimation of proper motion for the minor knots through use of either the observed or synthetic difference map would require each knot to have higher count statistics than is available in our epochs, hence why the cross-correlation method was preferred. Nonetheless, a qualitative comparison between the observed and synthetic difference maps may confirm the general features expected for our proper motion estimate. As no simulated noise was added to the synthetic map, the map appears less noisy than the observed map. Despite this minor difference, it is clear from the image comparison that the structures observed in the larger knots, such as AX1A, AX1C, and BX2, do not agree with those seen in the simulated map. In contrast, the simulated proper motion of 0.17\arcsec{} reproduces the overall structure observed in knots AX2, AX3, AX4, AX6, and BX1.

Thus, apart from the few brightest knots, the X-ray knot complexes A, B, and C within the inner jet of Cen\,A moved a total of $0.17 \pm 0.05\arcsec$ over a 15 yr timespan, giving an average proper motion of $11.3\pm3.3 \rm\ mas\ yr^{-1}$ over the length of the jet projected between 0.26 and 1.35 kpc from the AGN. This translates to a projected pattern speed of $\beta_{\rm app} = 0.68\pm0.20$. For a jet inclination of 50$^{\circ}$ \citep{Tingay1998, Hardcastle2003}, using the Doppler formula $\beta = \beta_{\rm app} / ({\rm sin} \theta + \beta_{ \rm app} {\rm cos} \theta)$, this gives a intrinsic pattern speed of $\beta = 0.57\pm0.11$, allowing only for the error in $\beta_{app}$.

%%%%%%%%%%%%%%%%%%%%%%%%%%%

\section{Discussion}
\label{sect:discuss}

%%%%%%%%%%%%%%%%%%%%%%%%%%%
\subsection{Comparison of Proper Motion to Other FR I Sources}
\label{sect:speed}

\begin{figure}
	\begin{tightcenter}
	\includegraphics[width=0.46\textwidth]{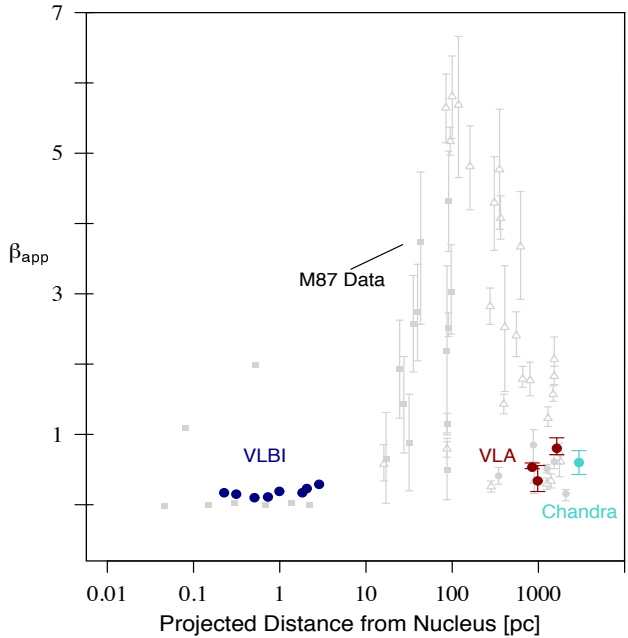}	
	\end{tightcenter} 		
    \caption{Comparison of the jet speeds versus distance in M87 (gray points) 
    		and Centaurus\,A. For Cen\,A, the VLBI points are taken from 
		\cite{Muller2014}, the VLA points from \cite{Goodger2010}, and 
		the \chandra{} data point from the current paper. The M87 data include 
		over a dozen studies with VLBI, the VLA, and {\it HST} 
		(adapted from \citealt{Meyer2013}, and references therein).}
	\label{fig:speed}
\end{figure}

Cen\,A is one of only three sources with measurements of the jet velocity on both parsec and kiloparsec scales. It is therefore interesting to compare the full velocity profile of Cen\,A to previously studied sources. By far the best-studied jet kinematically is that in M87, which has been studied with VLBI on scales from 0.01 to 100\,pc and with the VLA and the {\it Hubble Space Telescope} ({\it HST}) on scales from 100 to 1000\,pc (projected distances; \citealp{Meyer2013, Asada2014, Mertens2016, Walker2018}, and references therein). The acceleration zone in M87 beginning $<$~0.05\,pc (projected) from the nucleus appears to extend up to 100\,pc (projected), as tracked by increasing VLBI speeds, before an `explosion' at the stationary knot HST-1, which has been seen to emit components with apparent motions as fast as $6c$. Beyond this, gradual deceleration takes place (with slower and faster components), forming an ``envelope," as shown in Figure~\ref{fig:speed}. Here, we plot the historical M87 data (adapted from \citealp{Meyer2013}) in gray, and overlay the data points from Cen\,A in color. In particular, we plot the low VLBI speeds measured by \cite{Muller2014} in navy (error bars are smaller than the plotted points), and the previous VLA proper motions measured by \cite{Goodger2010} in red. Our \chandra{} data point (plotted at the mean distance of the boxed region used for the cross-correlation) is plotted in cyan.

It is notable that the speeds in Cen\,A are comparatively much lower than those seen in M87. They are also much slower than those seen in the only other source with kiloparsec-scale optical proper motions -- 3C\,264 -- which shows a remarkably similar envelope to M87 (though not as well sampled), including a peak speed of $\sim$\,$7c$ at approximately 100 pc from the core \citep{Meyer2015}. Here, we see that the span of the jet at which these high speeds emerge for M87 and 3C\,264 (roughly between ten and a few hundred pc) are not sampled by the observations of Cen\,A so far. It is therefore not completely clear if Cen\,A has a jet that intrinsically does not reach high (i.e., apparent superluminal) speeds, or if we simply miss the short part of the jet in which these speeds occur. Follow-up observations with moderately high-resolution interferometry, such as the Atacama Large Millimeter/submillimeter Array (ALMA) or Square Kilometer Array (SKA), could shed light on this matter. It should be noted that while 3C\,264 and M87 are both low-power FR\,Is ($10^{43.8}$ and $10^{43.7}$ erg\,s$^{-1}$ respectively; \citealp{Meyer2011}), Cen\,A has a jet power that is even lower (10$^{43.1}$ erg\,s$^{-1}$; \citealp{Croston2009, Wykes2013, Neff2015}). It is possible that the maximum speed is set by the jet power, and in this case that effect may be compounded by a larger angle to the line of sight in Cen\,A, compared to M87 or 3C\,264 \citep{Biretta1999, Giovannini2001}. 

%%%%%%%%%%%%%%%%%%%%%%%%%%%
\subsection{Synchrotron Cooling}
\label{sect:synch}

A possible cause for the observed decreases in brightness discussed in Section~\ref{sect:diffmap} is synchrotron cooling. For a given particle energy, the rate of cooling is governed by the magnetic field strength and the angle between a particle's velocity vector and the magnetic field, i.e., the pitch angle. For a fixed magnetic field strength, the cooling rate is maximized for particles moving perpendicular to the field. Under the Kardashev--Pacholczyk (KP) model \citep{Kardashev1962, Pacholczyk1970}, scattering is assumed to be negligible, so that the pitch angle only changes due to energy loss. Thus, the fastest possible cooling is for particles moving at $90^\circ$ to the field under the KP model. Assuming this model, we can estimate the minimum magnetic field strength required to cause the observed reductions in brightness of the statistically significant fading knots AX1C and BX2, both of which were found in Section~\ref{sect:motion}  to be stationary. For simplicity, all other sources of particle acceleration were considered to be negligible. Based on previously observed spectral indices for the X-ray knots \citep{Goodger2010}, we used an initial electron distribution of the form $dN/d\gamma = K \gamma^{-p}$, with a particle index $p = 3$ for AX1C and $p=2.2$ for BX2. A maximum Lorentz factor of $10^{9}$ was used in our estimates. The synchrotron spectrum was computed for the cooling electron distribution as a function of time, and the magnetic field strength was adjusted to produce the observed reduction in brightness at 1 keV. The brightness variations for the two knots, shown in Table~\ref{table:diffbright}, demonstrate a consistent decrease over the three epochs sampled, spanning 15 yr. We therefore assumed a total cooling timespan of 15 yr for our model. To obtain the fading reported for knot AX1C required a magnetic field strength of $\sim$\,80~$\mu$G, while knot BX2 also needed a magnetic field strength of $\sim$\,80~$\mu$G.

For a time-dependent synchrotron cooling model, it is almost inevitable that the spectrum steepens as the flux decreases. Defining the X-ray spectral index in terms of the flux as $F_E \sim E^{-\alpha}$ (photon flux $\sim E^{-(\alpha + 1)}$) for the KP model used, the observed brightness decreases would require the spectral index at 1 keV to evolve from $\alpha = 1$ initially to $\alpha \simeq 1.3$ for AX1C and from $\alpha = 0.6$ to $\alpha \simeq 0.8$ for knot BX2. To test for this effect, spectral indices were estimated for the knots in each epoch using a {\tt phabs\,$\times$\,powerlaw} model in \xspec{} 12.10.0c \citep{Arnaud1996}. The spectra were binned over 0.9--2.0 keV, where H{\sc I} column densities of 5.1$\times 10^{21}$ cm$^{-2}$ and 8.4 $\times$ 10$^{20}$ cm$^{-2}$ were used for AX1C and BX2, respectively, based on results from \cite{Dickey1990}. The observed photon indices from the 2002/2003 and 2017 epochs agreed within 1$\sigma$ for both AX1C ($1.98\pm0.09$ versus $1.93\pm0.08$, respectively) and BX2 ($1.40\pm0.08$ versus $1.45\pm0.09$, respectively). Given the discrepancy between the observed indices and those predicted from the synchrotron cooling model, we conclude that the decrease in brightness is unlikely to be from synchrotron cooling. The lack of evidence for the emission spectra steepening expected from synchrotron cooling makes it improbable that the actual magnetic field strengths exceed our estimates, so the value of 80~$\mu$G for both  AX1C and BX2 should therefore be viewed as an upper limit.

%%%%%%%%%%%%%%%%%%%%%%%%%%%
\subsection{Adiabatic Cooling}
\label{sect:adiabatic}

Instead of synchrotron cooling, adiabatic expansion may account for the observed rapid fading. Under the usual assumptions of isotropy, typical electron Lorentz factors scale under adiabatic expansion as $\gamma \sim V^{-1/3}$, with $V$ the volume, while the magnetic field strength scales as $B \sim V^{-2/3}$. For a power-law electron distribution of the form $dN/d\gamma \sim \gamma^{-p}$, the synchrotron flux at a fixed energy then scales as $F_E \sim V^{-2p/3}$. From this relationship, the observed changes in brightness could be used to place broad constraints on the geometry of the fading knots.

We initially considered one-dimensional expansion of the system where the knots are disk-like and all extension occurred along the line-of-sight at a speed of $c$ over 15 years. This approximation maximizes both the expansion rate and the diameter-to-width ratio of the disk. Using the changes in brightness from Table~\ref{table:diffbright}, the initial line-of-sight width was estimated as 59\,pc for AX1C and 85\,pc for BX2. The line-of-sight estimates were compared to the projected knot diameters, which were estimated from the 2002/2003 image using radial profile cuts from the knot centers as defined by the \ciao{} {\tt dmstat} routine. The knot diameter was measured as 40\,pc for AX1C and 140\,pc for BX2, giving a diameter-to-width ratio of 0.7 for AX1C and 1.6 for BX2. Even assuming the maximum expansion rate, the derived geometry is consistent with that expected for a front of shocked gas along the jet axis \citep{Komissarov1997}. Our approximation of one-dimensional adiabatic expansion is therefore physically consistent and agrees with the observations.

Although one-dimensional adiabatic expansion is plausible, it is more likely that the knots are undergoing complex expansion in a three-dimensional space. To approximate the average expansion rate for such a system, we considered the knots to be spherical where the volume expands equally in all directions. Using the changes in brightness from Table~\ref{table:diffbright} and the measured initial radii, the final knot radii after 15 yr of expansion were estimated to be 21\,pc for AX1C and 73\,pc for BX2. The rate of expansion equals $\sim$\,10 mas yr$^{-1}$, or $\sim$\,0.6$c$, for both knots, a change in the knots that is not resolvable in the difference maps due to the low count statistics at the outer knot edges. Despite lacking a clear detection of geometric changes indicative of expansion, we found the observations to be consistent with adiabatic expansion for both geometric cases. Since adiabatic expansion can be reconciled with observations for expansion in either one or three dimensions, it plausibly accounts for the dimming seen in the difference maps. This is consistent with radio observations of Cen\,A that have determined adiabatic expansion to be the primary cause of brightness variations \citep{Goodger2010}.

%%%%%%%%%%%%%%%%%%%%%%%%%%%
\subsection{Knot Origins}
\label{sect:knots}

We may constrain possible origins for the X-ray knots based on the \chandra{} observations. One knot formation mechanism is the short-term process of magnetic field reconnection, also referred to as impulsive particle acceleration. \cite{Goodger2010} pointed out that with an equipartition value of the magnetic field strength in the AX1A knot, $\sim$\,760~$\mu$G, one might see a complete change in the appearance of the X-ray emission in about six years. No such drastic changes are seen over our 15 yr timespan, and our magnetic field strength estimate of the adjacent knot AX1C (Section~\ref{sect:synch}) suggests a value no greater than $\sim$\,9 times lower than equipartition. This result supports particle acceleration processes that are longer-lived than reconnection.

Stationary knots, such as the A1 knots that were found to have a proper motion upper limit of $\sim$\,$0.1c$ (Section~\ref{sect:motion}), are frequently suggested to be associated with an obstacle in the jet. Non-destructive interactions of the jet plasma with gas or molecular clouds, such as the stellar winds of the $\sim$\,3 and $\sim$\,12 Gyr old populations known to reside in the elliptical galaxy \citep{Rejkuba2011}, remain a plausible proposition for the stationary knots. AGB stars of the 3 Gyr old population would certainly make for an effective obstruction, a topic that will be extensively discussed in a forthcoming paper \citep{Wykes2018a}. Here, we point to brightness changes seen in the upstream regions of knots AX1A and AX1C (Figure~\ref{fig:diffa}). Interestingly, the upstream part of AX1C appears to have faded in the 2017 images, while AX1A has brightened in its upstream region at the same epoch. Assuming that these knots are each associated with an obstacle in the jet, one would expect standing shocks in this region \citep{Hardcastle2003, Wykes2015, Wykes2018a}. These variations could arise from modest, local changes in flow speed, causing the bow shocks to vary in strength: a reduction in flow speed will cause the shock to weaken and advance upstream, while an increase in flow speed will have the opposite effect.
 
In contrast to the stationary knots, knots that are moving in the jet at velocities approaching the bulk flow speed are unlikely be a result of the jet plasma interacting with a relatively stationary obstacle. \cite{Goodger2010} postulated that a moderate compression of the jet fluid is the most likely scenario for the origin of the moving knots. Given that none of the observed X-ray knots extend over the whole diameter of the jet, such localized regions may be due to collisions of the jet plasma with an obstacle that is on the verge of being dissolved in the jet. This would apply to planetary nebulae and gas/molecular clouds. In addition to the knots, the existence of a well-defined peak in the cross-correlation at a significant offset from zero (Figure~\ref{fig:knot_cc}) requires a significant fraction of the fainter substructure within the jet to be moving at high speed in the jet's direction of flow. Were this due to changes in the jet power, one would expect to see features that cross the full jet. Since no such features are observed, the most likely cause is ``turbulence" in the jet, potentially generated by interaction with obstacles. This might cause compressions and weak shocks, or it might result in shear that modifies the magnetic field and/or accelerates particles. Follow-up deep-exposure X-rays observations could further clarify the interpretation of such substructure. Continued monitoring observations will also assist in constraining the overall lifetime of these features, and subsequently their physical nature.

%%%%%%%%%%%%%%%%%%%%%%%%%%%
\section{Conclusions}
\label{sect:conclusions}

Recent and archival \chandra{} observations were analyzed for evidence of variability and proper motion in the X-ray jet of Cen\,A. Data spanning 15 yr were co-aligned to high accuracy, and difference maps of the epochs were generated. Collective proper motion for the fainter substructure in the jet was measured by fitting the cross-correlation between epochs, finding a projected speed of $11.3\pm3.3 \rm\ mas\ yr^{-1}$ over the projected jet length 0.26--1.35~kpc from the AGN. This translates to a projected pattern speed of $\beta_{\rm app} = 0.68\pm0.20$, or an intrinsic pattern speed of $\beta = 0.57\pm0.11$, assuming a jet inclination angle of $50^{\circ}$. The cross-correlation results also imply that both the X-ray knots and a significant fraction of the substructure within the jet move at much the same speed, which is presumably the flow speed of the jet. Three of the brightest knots in the jet (AX1A, AX1C, BX2) were found to be stationary based on cross-correlation fits and centroid analyses, placing an upper limit of $\beta_{\rm app} < 0.10$ on each knot.  All measurements are consistent with proper motion estimates from VLA radio observations of knots within the same projected distance of Cen\,A. 

Proper motion estimates for Cen\,A were compared to jet velocity measurements for the other FR~I sources M87 and 3C\,264, and the speeds in Cen\,A were found to be comparatively lower. It remains to be determined whether this discrepancy may be attributed to the jet of Cen\,A not reaching such high speeds, differences in power and/or viewing angle between the various systems, or whether the small-scale region of the jet in which these high speeds happen has simply not been observed. Follow-up observations with higher -esolution instruments, such as ALMA or SKA, may help to resolve these issues.

Variations in brightness up to 27\% were detected for several X-ray knots in the jet, and potential mechanisms that may explain the observed fading of knots BX2 and AX1C were tested against the results. The changes in spectral slope expected to accompany fading due to synchrotron cooling were not found in the spectral analysis of knots BX2 and AX1C, ruling out synchrotron cooling and placing an upper limit of 80\,$\mu$G for both AX1C and BX2 on their magnetic field strengths. Adiabatic expansion was also tested as a potential fading mechanism for the knots, and it was found to be consistent with observations for expansion in one or three dimensions. Adiabatic expansion is therefore the mostly likely cause of the observed decreases in brightness of the knots. 

Our results were used to place constraints on models for the origin of the knots. Short-term acceleration processes, such as magnetic field reconnection, are disfavored based on the low magnetic field strength estimates and the lack of changes in appearance of the knots observed over the 15 yr timespan.  Stationary knots were best explained by the jet plasma overrunning an obstacle, stellar wind of a tip-AGB star or a cloud crossing the jet, with the stationary A1 knots showing potential evidence of standing shocks. In contrast, the moving knots may either arise from internal differences in jet speed or from late stages of the jet interaction with the nebular or cloud material. Deeper X-ray observations should help to define this substructure, while longer-term observations could constrain its lifetime, placing further constraints on its nature. 

\acknowledgements{
We thank the reviewer for his/her valuable comments and suggestions. Support for this work was provided by the National Aeronautics and Space Administration through {\it Chandra} Award Number G07-18104X issued by the {\it Chandra X-ray Observatory Center}, which is operated by the Smithsonian Astrophysical Observatory for and on behalf of the National Aeronautics Space Administration under contract NAS8-03060. S.W., P.E.J.N., and R.P.K. were supported in part by NASA contract NAS8-03060. S.W. thanks the Harvard-Smithsonian CfA for a research fellowship.
}

\software{
\ciao{} v4.9 \citep{Fruscione2006},
\xspec{} 12.10.0c \citep{Arnaud1996}
}

%%%%%%%%%%%%%%%%%%%%%%%%%%%%%%%%%%%%%%

\bibliographystyle{aasjournal}
\bibliography{all_data}

\begin{thebibliography}{}
\expandafter\ifx\csname natexlab\endcsname\relax\def\natexlab#1{#1}\fi
\providecommand{\url}[1]{\href{#1}{#1}}

\bibitem[{Al{\=u}zas {et~al.}(2012)Al{\=u}zas, Pittard, Hartquist, Falle, \&
  Langton}]{Aluzas2012}
Al{\=u}zas, R., Pittard, J.~M., Hartquist, T.~W., Falle, S. A. E.~G., \&
  Langton, R. 2012, MNRAS, 425, 2212

\bibitem[{Arnaud(1996)}]{Arnaud1996}
Arnaud, K.~A. 1996, in ASP Conference Series, Vol. 101, Astronomical Data
  Analysis Software and Systems V, ed. G.~H. Jacoby \& J.~Barnes

\bibitem[{Asada {et~al.}(2014)Asada, Nakamura, Doi, Nagai, \&
  Inoue}]{Asada2014}
Asada, K., Nakamura, M., Doi, A., Nagai, H., \& Inoue, M. 2014, ApJL, 781, L2

\bibitem[{Bicknell(1994)}]{Bicknell1994}
Bicknell, G.~V. 1994, ApJ, 422, 542

\bibitem[{Biretta {et~al.}(1999)Biretta, Sparks, \& Macchetto}]{Biretta1999}
Biretta, J.~A., Sparks, W.~B., \& Macchetto, F. 1999, ApJ, 520, 621

\bibitem[{Boccardi {et~al.}(2017)Boccardi, Krichbaum, Ros, \&
  Zensus}]{Boccardi2017}
Boccardi, B., Krichbaum, T.~P., Ros, E., \& Zensus, J.~A. 2017, Astron.
  Astrophys. Rev., 25, 4

\bibitem[{Breiding {et~al.}(2017)Breiding, Meyer, Georganopoulos, Keenan,
  DeNigris, \& Hewitt}]{Breiding2017}
Breiding, P., Meyer, E.~T., Georganopoulos, M., {et~al.} 2017, ApJ, 849, 95

\bibitem[{Canvin \& Laing(2004)}]{Canvin2004}
Canvin, J.~R., \& Laing, R.~A. 2004, MNRAS, 350, 1342

\bibitem[{Canvin {et~al.}(2005)Canvin, Laing, Bridle, \& Cotton}]{Canvin2005}
Canvin, J.~R., Laing, R.~A., Bridle, A.~H., \& Cotton, W.~D. 2005, MNRAS, 363,
  1223

\bibitem[{Cotton {et~al.}(1999)Cotton, Feretti, Giovannini, Lara, \&
  Venturi}]{Cotton1999}
Cotton, W.~D., Feretti, L., Giovannini, G., Lara, L., \& Venturi, T. 1999, ApJ,
  519, 108

\bibitem[{Croston {et~al.}(2009)Croston, Kraft, Hardcastle, Birkinshaw,
  Worrall, Nulsen, Penna, Sivakoff, Jordan, Brassington, Evans, Forman,
  Gilfanov, Goodger, Harris, Jones, Juett, Murray, Raychaudhury, Sarazin, Voss,
  \& Woodley}]{Croston2009}
Croston, J.~H., Kraft, R.~P., Hardcastle, M.~J., {et~al.} 2009, MNRAS, 395,
  1999

\bibitem[{Dickey \& Lockman(1990)}]{Dickey1990}
Dickey, J.~M., \& Lockman, F.~J. 1990, ARA\&A, 28, 215

\bibitem[{Fanaroff \& Riley(1974)}]{Fanaroff1974}
Fanaroff, B.~L., \& Riley, J.~M. 1974, MNRAS, 167, 31P

\bibitem[{Feigelson {et~al.}(1981)Feigelson, Schreier, Delvaille, Giacconi,
  Grindlay, \& Lightman}]{Feigelson1981}
Feigelson, E.~D., Schreier, E.~J., Delvaille, J.~P., {et~al.} 1981, ApJ, 251,
  31

\bibitem[{Fruscione {et~al.}(2006)Fruscione, McDowell, Allen, Brickhouse,
  Burke, Davis, Durham, Elvis, Galle, Harris, Huenemoerder, Houck, Ishibashi,
  Karovska, Nicastro, Nowak, Primini, Siemiginowska, Smith, \&
  Wise}]{Fruscione2006}
Fruscione, A., McDowell, J.~C., Allen, G.~E., {et~al.} 2006, Proc. SPIE, 6270,
  62701V

\bibitem[{Gentry {et~al.}(2015)Gentry, Marshall, Hardcastle, Perlman,
  Birkinshaw, Worrall, Lenc, Siemiginowska, \& Urry}]{Gentry2015}
Gentry, E.~S., Marshall, H.~L., Hardcastle, M.~J., {et~al.} 2015, ApJ, 808, 92

\bibitem[{Giovannini {et~al.}(2001)Giovannini, Cotton, Feretti, Lara, \&
  Venturi}]{Giovannini2001}
Giovannini, G., Cotton, W.~D., Feretti, L., Lara, L., \& Venturi, T. 2001, ApJ,
  552, 508

\bibitem[{Goodger {et~al.}(2010)Goodger, Hardcastle, Croston, Kraft,
  Birkinshaw, Evans, Jord{\'a}n, Nulsen, Sivakoff, Worrall, Brassington,
  Forman, Gilfanov, Jones, Murray, Raychaudhury, Sarazin, Voss, \&
  Woodley}]{Goodger2010}
Goodger, J.~L., Hardcastle, M.~J., Croston, J.~H., {et~al.} 2010, ApJ, 708, 675

\bibitem[{Hada {et~al.}(2016)Hada, Kino, Doi, Nagai, Honma, Akiyama, Tazaki,
  Lico, Giroletti, Giovannini, Orienti, \& Hagiwara}]{Hada2016}
Hada, K., Kino, M., Doi, A., {et~al.} 2016, ApJ, 817, 131

\bibitem[{Hardcastle {et~al.}(2001)Hardcastle, Birkinshaw, \&
  Worrall}]{Hardcastle2001}
Hardcastle, M.~J., Birkinshaw, M., \& Worrall, D.~M. 2001, MNRAS, 326, 1499

\bibitem[{Hardcastle {et~al.}(2006)Hardcastle, Kraft, \&
  Worrall}]{Hardcastle2006}
Hardcastle, M.~J., Kraft, R.~P., \& Worrall, D.~M. 2006, MNRAS, 368, L15

\bibitem[{Hardcastle {et~al.}(2003)Hardcastle, Worrall, Kraft, Forman, Jones,
  \& Murray}]{Hardcastle2003}
Hardcastle, M.~J., Worrall, D.~M., Kraft, R.~P., {et~al.} 2003, ApJ, 593, 169

\bibitem[{Hardcastle {et~al.}(2007)Hardcastle, Kraft, Sivakoff, Goodger,
  Croston, Jord{\'a}n, Evans, Worrall, Birkinshaw, Raychaudhury, Brassington,
  Forman, Harris, Jones, Juett, Murray, Nulsen, Sarazin, \&
  Woodley}]{Hardcastle2007}
Hardcastle, M.~J., Kraft, R.~P., Sivakoff, G.~R., {et~al.} 2007, ApJL, 670, L81

\bibitem[{Harris {et~al.}(1997)Harris, Biretta, \& Junor}]{Harris1997}
Harris, D.~E., Biretta, J.~A., \& Junor, W. 1997, MNRAS, 284, L21

\bibitem[{Harris {et~al.}(2010)Harris, Rejkuba, \& Harris}]{Harris2010}
Harris, G. L.~H., Rejkuba, M., \& Harris, W.~E. 2010, Publ. Astron. Soc. Aust.,
  27, 457

\bibitem[{Kardashev(1962)}]{Kardashev1962}
Kardashev, N.~S. 1962, SvA, 6, 317

\bibitem[{Kataoka {et~al.}(2006)Kataoka, Stawarz, Aharonian, Takahara,
  Ostrowski, \& Edwards}]{Kataoka2006}
Kataoka, J., Stawarz, L., Aharonian, F., {et~al.} 2006, ApJ, 641, 158

\bibitem[{Komissarov \& Falle(1997)}]{Komissarov1997}
Komissarov, S.~S., \& Falle, S. A. E.~G. 1997, MNRAS, 288, 833

\bibitem[{Kraft {et~al.}(2002)Kraft, Forman, Jones, Murray, Hardcastle, \&
  Worrall}]{Kraft2002}
Kraft, R.~P., Forman, W.~R., Jones, C., {et~al.} 2002, ApJ, 569, 54

\bibitem[{Laing \& Bridle(2002)}]{Laing2002}
Laing, R.~A., \& Bridle, A.~H. 2002, MNRAS, 336, 1161

\bibitem[{Laing {et~al.}(1999)Laing, Parma, de~Ruiter, \& Fanti}]{Laing1999}
Laing, R.~A., Parma, P., de~Ruiter, H.~R., \& Fanti, R. 1999, MNRAS, 306, 513

\bibitem[{Lister {et~al.}(2013)Lister, Aller, Aller, Homan, Kellermann,
  Kovalev, Pushkarev, Richards, Ros, \& Savolainen}]{Lister2013}
Lister, M.~L., Aller, M.~F., Aller, H.~D., {et~al.} 2013, AJ, 146, 120

\bibitem[{Marshall {et~al.}(2002)Marshall, Miller, Davis, Perlman, Wise,
  Canizares, \& Harris}]{Marshall2002}
Marshall, H.~L., Miller, B.~P., Davis, D.~S., {et~al.} 2002, ApJ, 564, 683

\bibitem[{Mertens {et~al.}(2016)Mertens, Lobanov, Walker, \&
  Hardee}]{Mertens2016}
Mertens, F., Lobanov, A. P.~., Walker, R.~C., \& Hardee, P.~E. 2016, A\&A, 595,
  A54

\bibitem[{Meyer {et~al.}(2011)Meyer, Fossati, Georganopoulos, \&
  Lister}]{Meyer2011}
Meyer, E.~T., Fossati, G., Georganopoulos, M., \& Lister, M.~L. 2011, ApJ, 740,
  98

\bibitem[{Meyer \& Georganopoulos(2014)}]{Meyer2014}
Meyer, E.~T., \& Georganopoulos, M. 2014, ApJL, 780, L27

\bibitem[{Meyer {et~al.}(2015)Meyer, Georganopoulos, Sparks, Godfrey, Lovell,
  \& Perlman}]{Meyer2015}
Meyer, E.~T., Georganopoulos, M., Sparks, W.~B., {et~al.} 2015, ApJ, 805, 154

\bibitem[{Meyer {et~al.}(2013)Meyer, Sparks, Biretta, Anderson, Sohn, van~der
  Marel, Norman, \& Nakamura}]{Meyer2013}
Meyer, E.~T., Sparks, W.~B., Biretta, J.~A., {et~al.} 2013, ApJL, 774, L21

\bibitem[{Meyer {et~al.}(2017)Meyer, Sparks, Georganopoulos, van~der Marel,
  Anderson, Sohn, Biretta, Norman, Chiaberge, \& Perlman}]{Meyer2017}
Meyer, E.~T., Sparks, W.~B., Georganopoulos, M., {et~al.} 2017, Galaxies, 5,
  154

\bibitem[{M{\"u}ller {et~al.}(2014)M{\"u}ller, Kadler, Ojha, Perucho,
  Gro{\ss}Ÿberger, Ros, Wilms, Blanchard, B{\"o}ck, Carpenter, Dutka, Edwards,
  Hase, Horiuchi, Kreikenbohm, Lovell, Markowitz, Phillips, Pl{\"o}tz, Pursimo,
  Quick, Rothschild, Schulz, Steinbring, Stevens, Tr{\"u}stedt, \&
  Tzioumis}]{Muller2014}
M{\"u}ller, C., Kadler, M., Ojha, R., {et~al.} 2014, A\&A, 569, A115

\bibitem[{Nagai {et~al.}(2014)Nagai, Haga, Giovannini, Doi, Orienti, D'Ammando,
  Kino, Nakamura, Asada, Hada, \& Giroletti}]{Nagai2014}
Nagai, H., Haga, T., Giovannini, G., {et~al.} 2014, ApJ, 785, 53

\bibitem[{Neff {et~al.}(2015)Neff, Eilek, \& Owen}]{Neff2015}
Neff, S.~G., Eilek, J.~A., \& Owen, F.~N. 2015, ApJ, 802, 88

\bibitem[{Pacholczyk(1970)}]{Pacholczyk1970}
Pacholczyk, A.~G. 1970, Radio astrophysics. Nonthermal processes in galactic
  and extragalactic sources (San Franciso: W. H. Freeman)

\bibitem[{Perucho {et~al.}(2014)Perucho, Mart{\'i}­, Laing, \&
  Hardee}]{Perucho2014b}
Perucho, M., Mart{\'i}­, J.~M., Laing, R.~A., \& Hardee, P.~E. 2014, MNRAS,
  441, 1488

\bibitem[{Peterson {et~al.}(1998)Peterson, Wanders, Horne, Collier, Alexander,
  Kaspi, \& Maoz}]{Peterson1998}
Peterson, B., Wanders, I., Horne, K., {et~al.} 1998, Publ. Astron. Soc. Pac.,
  110, 660

\bibitem[{Rejkuba {et~al.}(2011)Rejkuba, Harris, Greggio, \&
  Harris}]{Rejkuba2011}
Rejkuba, M., Harris, W.~E., Greggio, L., \& Harris, G. L.~H. 2011, A\&A, 526,
  A123

\bibitem[{Snios {et~al.}(2018)Snios, Nulsen, Wise, de~Vries, Birkinshaw,
  Worrall, Duffy, Kraft, McNamara, Carilli, Croston, Edge, Godfrey, Hardcastle,
  Harris, Laing, Mathews, McKean, Perley, Rafferty, \& Young}]{Snios2018b}
Snios, B., Nulsen, P. E.~J., Wise, M.~W., {et~al.} 2018, ApJ, 855, 71

\bibitem[{Tingay \& Lenc(2009)}]{Tingay2009}
Tingay, S.~J., \& Lenc, E. 2009, AJ, 138, 808

\bibitem[{Tingay {et~al.}(1998)Tingay, Jauncey, Reynolds, Tzioumis, King,
  Preston, Jones, Murphy, Meier, van Ommen, McCulloch, Ellingsen, Costa,
  Edwards, Lovell, Nicolson, Quick, Kemball, Migenes, Harbison, Jones, White,
  Gough, Ferris, Sinclair, \& Clay}]{Tingay1998}
Tingay, S.~J., Jauncey, D.~L., Reynolds, J.~E., {et~al.} 1998, AJ, 115, 960

\bibitem[{Turner {et~al.}(1997)Turner, George, Mushotzky, \&
  Nandra}]{Turner1997}
Turner, T.~J., George, I.~M., Mushotzky, R.~F., \& Nandra, K. 1997, ApJ, 475,
  118

\bibitem[{Walker {et~al.}(2018)Walker, Hardee, Davies, Ly, \&
  Junor}]{Walker2018}
Walker, R.~C., Hardee, P.~E., Davies, F.~B., Ly, C., \& Junor, W. 2018, ApJ,
  855, 128

\bibitem[{Worrall {et~al.}(2001)Worrall, Birkinshaw, \&
  Hardcastle}]{Worrall2001}
Worrall, D.~M., Birkinshaw, M., \& Hardcastle, M.~J. 2001, MNRAS, 326, L7

\bibitem[{Worrall {et~al.}(2010)Worrall, Birkinshaw, O'Sullivan, Zezas, Wolter,
  Trinchieri, \& Fabbiano}]{Worrall2010}
Worrall, D.~M., Birkinshaw, M., O'Sullivan, E., {et~al.} 2010, MNRAS, 408, 701

\bibitem[{Worrall {et~al.}(2008)Worrall, Birkinshaw, Kraft, Sivakoff,
  Jord\'{a}n, Hardcastle, Brassington, Croston, Evans, Forman, Harris, Jones,
  Juett, Murray, Nulsen, Raychaudhury, Sarazin, \& Woodley}]{Worrall2008}
Worrall, D.~M., Birkinshaw, M., Kraft, R.~P., {et~al.} 2008, ApJL, 673, L135

\bibitem[{Wykes {et~al.}(2015)Wykes, Hardcastle, Karakas, \& Vink}]{Wykes2015}
Wykes, S., Hardcastle, M.~J., Karakas, A.~I., \& Vink, J.~S. 2015, MNRAS, 447,
  1001

\bibitem[{Wykes {et~al.}(2018{\natexlab{a}})Wykes, Snios, Nulsen, Kraft,
  Hardcastle, \& more}]{Wykes2018b}
Wykes, S., Snios, B., Nulsen, P. E.~J., {et~al.} 2018{\natexlab{a}}, MNRAS

\bibitem[{Wykes {et~al.}(2013)Wykes, Croston, Hardcastle, Eilek, Biermann,
  Achterberg, Bray, Lazarian, Haverkorn, Protheroe, \& Bromberg}]{Wykes2013}
Wykes, S., Croston, J.~H., Hardcastle, M.~J., {et~al.} 2013, A\&A, 558, A19

\bibitem[{Wykes {et~al.}(2018{\natexlab{b}})Wykes, Perucho, Jones, McDonald,
  Hardcastle, Karakas, Rejkuba, Nulsen, Bosch-Ramon, Kraft, Ferrand, Henkel,
  Richardson, Zijlstra, O?Dea, Vink, Safi-Harb, \& Baum}]{Wykes2018a}
Wykes, S., Perucho, M., Jones, T.~W., {et~al.} 2018{\natexlab{b}}, MNRAS

\end{thebibliography}

\end{document}